\documentclass[rotate]{emulateapj}

\newcommand{\lya}{Ly-$\alpha$~}
\newcommand{\lyb}{Ly-$\beta$~}

\newcommand{\nhi}{N_{\mhi}}

\newcommand{\cii}{C~{\sc ii}~}
\newcommand{\ciii}{C~{\sc iii}~}
\newcommand{\civ}{C~{\sc iv}~}

\newcommand{\ovi}{O~{\sc vi}~}

\newcommand{\siii}{Si~{\sc ii}~}
\newcommand{\siiii}{Si~{\sc iii}~}
\newcommand{\siiv}{Si~{\sc iv}~}
\newcommand{\hi}{H~{\sc i}~}

\newcommand{\heii}{He~{\sc ii}~}
\newcommand{\nv}{N~{\sc v}~}
\newcommand{\feii}{Fe~{\sc ii}~}
\newcommand{\alii}{Al~{\sc ii}~}

\newcommand{\nii}{N~{\sc ii}~}
\newcommand{\mgii}{Mg~{\sc ii}~}
\newcommand{\oden}{\rho/\bar{\rho}}

\newcommand{\mhi}{{\rm H \; \mbox{\tiny I}}}

\newcommand{\movi}{{\rm O \; \mbox{\tiny VI}}}

\newcommand{\mciv}{{\rm C \; \mbox{\tiny IV}}}
\newcommand{\msiiv}{{\rm Si \; \mbox{\tiny IV}}}
\newcommand{\msiii}{{\rm Si \; \mbox{\tiny II}}}
\newcommand{\mcii}{{\rm C \; \mbox{\tiny II}}}
\newcommand{\mciii}{{\rm C \; \mbox{\tiny III}}}
\newcommand{\msiiii}{{\rm Si \; \mbox{\tiny III}}}
\newcommand{\mnii}{{\rm N \; \mbox{\tiny II}}}
\newcommand{\mmgii}{{\rm Mg \; \mbox{\tiny II}}}
\newcommand{\mnv}{{\rm N \; \mbox{\tiny V}}}

\newcommand{\malii}{{\rm Al \; \mbox{\tiny II}}}
\newcommand{\mfeii}{{\rm Fe \; \mbox{\tiny II}}}
\newcommand{\dagnote}{\tablenotemark{$\dagger$}}

\newcommand{\hinv}{h_{71}^{-1}}
\newcommand{\ha}{H$\alpha$ ~}
\newcommand{\kms}{km s$^{-1}$}

\begin{document}
 
\bibliographystyle{/h0/simcoe/latex/apj}
 
%
\title{Observations of Chemically Enriched QSO Absorbers near $z\sim
2.3$ Galaxies: Galaxy-Formation Feedback Signatures in the
IGM\altaffilmark{1}}
 
\author{Robert A. Simcoe\altaffilmark{2,3}, Wallace
L.W. Sargent\altaffilmark{4}, Michael Rauch\altaffilmark{5}, George Becker\altaffilmark{4}}
 
\altaffiltext{1}{Includes observations made at the W.M. Keck
Observatory, which is operated as a scientific partnership between the
California Institute of Technology and the University of California;
it was made possible by the generous support of the W.M. Keck
Foundation.}  \altaffiltext{2}{MIT Center for Space Research, 77
Massachusetts Ave.  \#37-664B, Cambridge, MA 02139, USA;
simcoe@mit.edu} \altaffiltext{3}{Pappalardo Fellow in Physics}
\altaffiltext{4}{Palomar Observatory, California Institute of
Technology, Pasadena, CA 91125, USA; wws@astro.caltech.edu,
gdb@astro.caltech.edu} \altaffiltext{5}{Carnegie Observatories, 813
Santa Barbara Street, Pasadena, CA 91101, USA; mr@ociw.edu}
 
\begin{abstract}

We present a comparative study of galaxies and intergalactic gas
toward the $z=2.73$ quasar HS1700+6416, to explore the effects of
galaxy formation feedback on the IGM.  Our observations and ionization
simulations indicate that the volume within $100-200\hinv$ physical
kpc of high-redshift galaxies is populated by very small ($\Delta
L\lesssim 1$ kpc), dense ($\oden \sim 1000$), and metal-rich
($Z\gtrsim \frac{1}{10}-\frac{1}{3} Z_\sun$) absorption-line regions.
These systems often contain shock-heated gas seen in \ovi, and may
exhibit [Si/C] abundance enhancements suggestive of preferential
enrichment by Type II supernovae.  We argue that the absorber
geometries resemble thin sheets or bubbles, and that their unusual
physical properties can be explained using a simple model of
radiatively efficient shocks plowing through moderately overdense
intergalactic filaments.  The high metallicities suggest that these
shocks are being expelled from---rather than falling into---star
forming galaxies.  There is a dropoff in the intergalactic gas density
at galaxy impact parameters of $\gtrsim 300$ physical kpc ($\gtrsim 1$
comoving Mpc) that may represent boundaries of the gas structures
where galaxies reside.  The heavy-element enhancement near galaxies
covers smaller distances: at galactocentric radii between
$100-200\hinv$ kpc the observed abundances blend into the general
metallicity field of the IGM.  Our results suggest that either
supernova-driven winds or dynamical stripping of interstellar gas
alters the IGM near massive galaxies, even at $R\gtrsim 100$ kpc.
However, only a few percent of the total mass in the \lya forest is
encompassed by this active feedback at $z\sim 2.5$.  The effects
could be more widespread if the more numerous metal-poor \civ systems
at impact parameters $\gtrsim 200\hinv$ kpc are the tepid remnants of
very powerful late-time winds.  However, based on present observations
it is not clear that this scenario is to be favored over one involving
pre-enrichment by smaller galaxies at $z\gtrsim 6$.

\end{abstract}


\section{Introduction}\label{sec_introduction}

A wealth of evidence suggests that early galaxies expelled chemically
enriched gas widely into their surroundings.  This material may have
been thrown off by shocks or tidal interactions as proto-galactic
clumps collided, formed stars, and merged into larger units
\citep{gnedin_mergers, merger_feedback}.  Or, the feedback may have
been powered by supernovae in the star forming regions themselves
\citep[e.g.,][]{maclow_mccray}.  Whatever the mechanism, these early
galaxy/IGM interactions must have been much more vigorous than we
observe in the present day.

Starburst galaxies are often accompanied by supernova-driven outflows
which can carry enriched material over intergalactic distances
\citep{starburst_review}.  At low redshift, these so-called superwinds
preferentially occur in dwarf galaxies whose halo escape velocities
are small \citep[e.g.,][]{heckman_ngc1705,crystal_wind_yield}.  At
high redshift, the star formation density was much higher, so the
corresponding increase in supernova frequency should allow larger
galaxies to drive superwinds.  This phenomenon is indeed seen in
$z\gtrsim 3$ galaxies, most of which bear the the spectroscopic
signature of metal-rich outflows \citep{franx_winds, lbg_winds}.

Ultimately, any feedback theory must explain how the IGM was populated
with heavy elements from very early times.  No decline is observed in
the \civ contribution to closure density out to redshift $z\gtrsim 5$
\citep{pettini_z5_civ,songaila_omegaz}.  Moreover by $z\sim 2-3$, \civ
and \ovi can be observed in gas with density near the cosmic mean
\citep{simcoe2004,schaye_civ_pixels, aguirre_siiv_pixels}.  From the
observed abundances, one infers that within the first $15\%$ of the
Hubble time, $\sim 50\%$ of all baryons (i.e., most of the cosmic
filaments) were mixed with chemically-rich gas.  To produce the
observed abundances, the average galaxy at $z\gtrsim 2.5$ would need
to eject $\gtrsim 15\%$ of its manufactured metals into the IGM
\citep{simcoe2004}.

A relationship between high-redshift galactic winds and the observed
intergalactic abundances has long been postulated, yet relatively few
details are known about how the winds' chemicals and energy physically
mix over large scales.  \citet{kurt_winds, adelberger_z2} have studied
the large-scale cross-correlation between $z\sim 2-3$ Lyman break
galaxies, \lya forest lines, and \civ systems.  They found a
positive correlation between galaxies and \civ systems extending over
$4.2\hinv$ comoving Mpc.  Moreover the strongest \civ absorbers
correlate more strongly with galaxies than galaxies do with each
other.  This suggests that these ``strong'' \civ systems (roughly
$N_\mciv\gtrsim 10^{13}$) congregate directly around individual galaxy
haloes.  They interpret the galaxy-\civ clustering as a signature of
late superwinds, but similar analyses at low redshift
\citep{chen_civ_galaxies} and at high redshift using quasar spectra
alone \citep{rauch_simulations} find that the same observations can be
explained by accretion of pre-enriched gas during hierarchical galaxy
assembly \citep[see also,][]{porciani_madau_lbg_metals}.

At $z\sim 2.5$, high column density \civ systems are often accompanied
by strong \ovi absorption, as would be found in hot ($T\gtrsim 10^5$),
shock-heated environments \citep{simcoe2002}.  Typically these
absorbers are marginally optically thin ($\nhi\sim
10^{15}-10^{16.5}$), but they display a very rich chemical structure,
containing many heavy elements in a broad range of temperatures and
ionization states.  The strong assocation between \civ systems and
galaxies at higher redshift, the presence of shock-heating, and the
apparently strong metal enrichment together suggest that the strongest
\ovi and \civ systems trace late stages of galaxy feedback, where
chemicals and energy mix into the IGM in real-time.  Very recent
reports of a strong correlation between galaxies and \ovi aborption
lend further support to this hypothesis \citep{adelberger_z2}.

In this paper, we dissect the physical properties of six metal-rich
absorption line systems, and examine their relationship with nearby
star-forming galaxies.  The systems are primarily selected for \ovi
absorption, but supplemented by one \nv and one \mgii identification.
Redshifts near $z\sim 2.5$ are particularly convenient for these
investigations, from the perspective of both the absorbers and the
galaxies.  A host of important rest-frame UV lines which vary widely
in ionization potential can be observed at optical wavelengths from
the ground.  This enables us to study intergalactic absorption systems
in physical states ranging from hot and collisionally ionized (\ovi,
\nv), to warm and photoionized (\siiv, \civ), to cool and nearly
neutral (\feii, \mgii, \alii).  Also, broadband color-selection
techniques are efficient at isolating $z\sim 2.5$ galaxies
\citep{steidel_desert}.  Galaxy samples reaching the equivalent of
$L^*$ and below may be assembled using reasonable multislit
integrations with 6-10m class telescopes.

Our preliminary investigations are based upon six absorption systems
and 14 galaxies towards a single quasar sightline.  We construct
detailed component-by-component ionization models of the absorbers,
using their full complement of heavy-element transitions to constrain
gas densities, metallicities, [Si/C] abundance enhancements, sizes,
and temperatures.  These properties are examined in light of each
systems' stellar neighborhood, which is charted using color selected,
spectroscopically confirmed high redshift galaxies.  In Section
\ref{sec:observations} we describe the observational program, followed
by a discussion of our absorption line fits and ionization simulations
in Sections \ref{sec:vpfit} and \ref{sec:analysis}.  In Section
\ref{sec:physical_model} we demonstrate how the absorption line
properties are naturally explained by a generic model of a radiative
shocks propagating into cosmic filaments.  Finally, in Section
\ref{sec:discussion} we elaborate on the connection between
high-metallicity absorbers and nearby galaxies, and comment on the
significance of galaxy feedback in a cosmological context.

\section{Data and Observations}\label{sec:observations}

Our observations are centered upon the $z=2.73$ QSO HS1700+6416
\citep{reimers_Q1700_discovery}.  We first observed HS1700 with HIRES
in April 2001 as part of a multi-sightline survey for intergalactic
\ovi absorption \citep{simcoe2002}.

HS1700 is very bright ($V\sim 16.1$) and it is one of only a few
objects suitable for studying the \heii \lya forest at $z\sim
2.5-3.0$.  Accordingly, it has been observed extensively from the
ground \citep{tripp_HS1700}, and also from space using FOS
\citep{HS1700_FOS} and FUSE \citep{FUSE_HS1700}.  These authors have
all noted the presence of several near-Lyman-limit systems along the
line of sight, which are optically thin yet exhibit high apparent
metallicities.  To investigate these systems in greater detail, we
have obtained high signal-to-noise ratio HIRES spectra with coverage
from $3200$\AA~ to nearly $1 \mu$m.  This enables measurements of a
large host of ion transitions for systems in the $1.8 \lesssim z
\lesssim 2.8$ range.  In the full sample, we have observed lines from
\hi, \cii, \ciii, \civ, \siii, \siiii, \siiv, \ovi, \nii, \nv, \alii,
\mgii, and \feii.  This list covers a wide range of ionization
potentials, providing considerable leverage for our ionization models.

\begin{figure}
\plotone{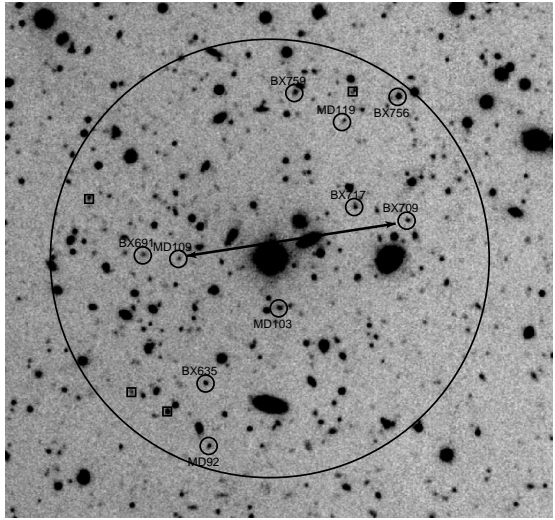}
\caption{$G$ band image of HS1700+6416 field.  Spectroscopically
identified $z\sim 1.8-2.7$ galaxies are labeled with circles (See
Table 1 for galaxy properties).  Unconfirmed photometric candidates
are indicated with square boxes.  The large circle indicates an impact
parameter of 500$\hinv$ physical kpc from the QSO line of sight at
$z=2.4$; the radial limit of the sample.  MD109 and BX709 have very
similar redshifts ($z=2.285,2.2942$) and may be related to a
single-component \civ system at $z=2.28956$ (See Section
\ref{sec:q1700_cluster}).  The faintest objects in the image have
$G\approx 28$; accordingly the $R<25.5$ sample limit is dictated by
spectrscopic rather than photometric sensitivities.  Image provided
courtesy of C. Steidel.}
\label{fig:q1700_field}
\end{figure}

\subsection{Galaxy Observations}\label{sec:galaxy_obs}

Recently the HS1700 field has also been targeted as part of a $z\sim
1.4-2.5$ galaxy survey by \citet{steidel_desert}.  
The authors compiled lists of candidate objects from deep, wide-field
$U_n,G,R$ broadband images, using the ``BX'' and ``MD'' color
selection criteria outlined in \citet{adelberger_selection}.  They
obtained spectroscopic redshifts for $\sim 100$ candidates using
Keck/LRIS-B.  These galaxies' colors and redshifts were kindly
provided to us by C. Steidel.  We focus on a small subset of the total
galaxy sample, containing objects with impact parameters $\rho <
500h_{71}^{-1}$ kpc (physical) from the QSO sightline
\footnote{When calculating angular size and luminosity distances, we
  assume a flat cosmology throughout, with $\Omega_{M}=0.3,
  \Omega_{\Lambda}=0.7, H_0=71$}.  

Fourteen photometric candidates meet this criterion, and their
properties are summarized in Table 1.  Ten of these galaxies have been
observed successfully for redshifts, three have been attempted
unsuccessfully, and one has not been attempted.  The redshift
identifications are 100\% complete for photometric candidates at
$R<25.5$, within $\sim 400 h_{71}^{-1}$ physical kpc of the QSO
sightline.  However, the photometric parent sample is probably only
$\sim 50-60\%$ complete at these redshifts because of object blending
or noise, or because some $z\sim 2.3$ galaxies fall outside of the
color selection boundaries (C. Steidel, private communication).

Figure \ref{fig:q1700_field} presents a summary image of the field,
with the relative locations of the QSO and $z\sim 2.5$ galaxies.  The
large circle shows the approximate $500\hinv$ kpc sample boundary;
small circles indicate objects with confirmed redshifts, and
unconfirmed photometric candidates are indicated with squares.  We are
primarily interested in foreground galaxy/absorber systems, so we have
omitted galaxies with $z\ge2.7$, which are close to the QSO's emission
redshift ($z_{QSO}=2.73$).  The galaxy statistics in this environment
could be biased by clustering around the QSO, and the absorption
systems would be subject to a locally anomolous radiation field.

The photometric limit of the candidate sample is $R<25.5$,
corresponding to $M_{\rm 2000\AA}=-20.8$ at $z\sim 2.3$.  It is not
entirely straightforward to relate this to a luminosity function due
to evolutionary and bandpass effects.  However, to provide some
context we note that for $R$ band observations of $z\sim 3$ galaxies
(corresponding to $1700$\AA ~in the rest-frame), $m^{*}_{(z=3)}=24.48$
\citep{steidel_lf}.  At $z\sim 2.3$, this translates to
$m^{*}_{(z=2.3)}=24.0$ at $\lambda_{\rm rest}\approx 2000$\AA ~using a
naive scaling which accounts simply for luminosity distance.  By this
measure the luminosity cutoff of the HS1700 galaxy sample is $\sim
\frac{1}{4}L^*$. The true cutoff could be slightly brighter due to
downward evolution of the luminosity function between $z\sim3$ and
$z\sim 2.5$ \citep[$\Delta m\sim 0.2-0.3$,][]{galex_lf_evolution} or
color differences from $1700$\AA ~to $2000$\AA.  However, early
compilations of galaxy samples in our redshift range find a nearly
identical luminosity function at $z\sim 3$ and $z\sim 2.2$
\citep{reddy_z2_lf}.

\citet{erb_halpha} have also obtained $K$ band spectra of the four
sample galaxies nearest the quasar sightline (indicated in Table 1).
These data cover the \ha transition, whose emission line is a much
more reliable estimate of the galaxies' stellar redshift than
rest-frame UV lines (which are often offset from the true systemic
redshift).  Much of our analysis concerns these closest systems, and
we adopt Erb's \ha redshifts where possible.  We also quote the star
formation rates derived from their \ha observations.  Finally, Table 1
also lists stellar mass estimates for each galaxy, based upon stellar
population models that incorporate rest-frame UV through near-IR
photometry \citep{alice_spitzer}.

\subsection{HIRES Observations}

We observed HS1700+6416 with HIRES on several different occasions,
using different instrument configurations to capture transitions
ranging from the very blue (\ovi, \nv) to the very red (\feii, \mgii).
All spectra were obtained through an $0.86$ arcsecond slit, which
corresponds to a velocity resolution of $\Delta v=6.6$ km/s (FWHM).
We extracted 2-D echelle spectra using T. Barlow's MAKEE reduction
package, fit cubic spline continua to each echelle order, and divided
the data by the model continuum.  Finally, the unity normalized
spectra were rebinned onto constant velocity pixels and the orders
were combined to produce a single, nearly continuous spectrum ranging
from $3200\AA-1\mu$m (there is a coverage gap in the $6200-7300$\AA
~range which is sparsely populated by $z\sim 2.5$ ions).  Our
continuum determination should be quite accurate in regions redward of
the QSO's \lya emission line, with errors probably below $\lesssim
0.5-1\%$ because of the data's high signal to noise.  Within the \lya
forest the errors are difficult to estimate.  There are still numerous
portions of clean spectrum that could be used to constrain the shape
of the spline, but in some heavily blended regions the errors could
possibly amount to $5-10\%$.

We searched the 1-D spectrum visually for highly ionized absorption
systems, which were identified via \ovi and/or \nv doublets.  Both
these transitions are blended amongst \hi lines from the \lya and \lyb
forests, which considerably complicates their identification.  To
improve our success rate, we adopted the procedure of
\citet{carswell2002}, fitting absorption profiles to the entire \lya
forest and using these parameters to remove the absorption signatures
of the corresponding \lyb and higher order lines.  A search of the
``cleaned'' spectrum for \ovi and \nv absorption yielded five clear
detctions---four identified from \ovi, and one from \nv.  The one
system identified via \nv is located at $z=1.84$, where \ovi cannot be
observed from the ground due to the atmospheric UV cutoff.

In addition to these five systems, we found a sixth example that
appeared to be quite metal rich, based upon the detection of \civ,
\cii, \siii, \siiii, \siiv, and \mgii.  All six systems are
accompanied by \civ absorption with $N_\mciv>10^{13}$, which puts them
in league with the strong \civ systems that cluster strikingly with
galaxies at $z\sim 3$ \citep{kurt_winds, adelberger_z2}.  They also
exhibit strong, sub-Lyman-limit \hi absorption ($\nhi\sim
10^{15-16}$), and a rich assortment of lower ionization potential
heavy element lines.

\section{Absorption Line Characterization}\label{sec:vpfit}

\begin{figure}
\plotone{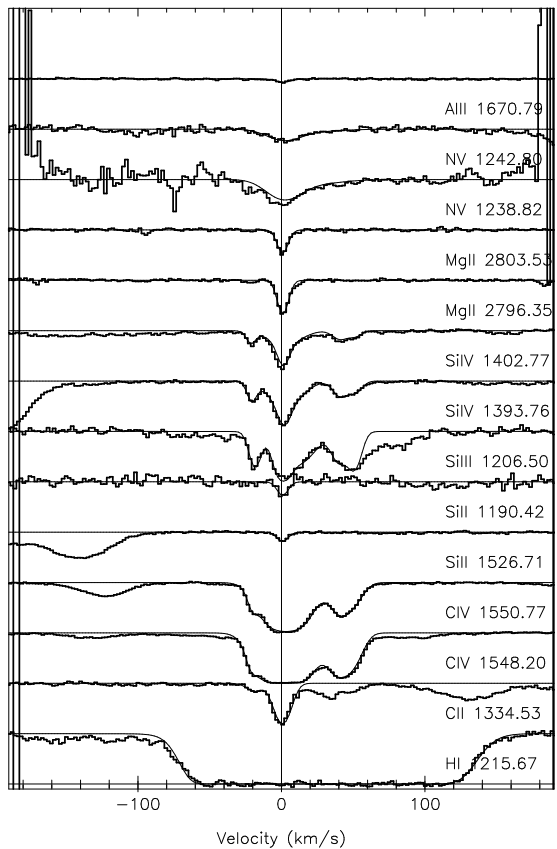}
\caption{Stacked velocity plot of heavy element absorption lines at
$z=1.846$.  See Section \ref{sec:z1.846} for description}
\label{fig:z1.84}
\end{figure}

\begin{figure}
\plotone{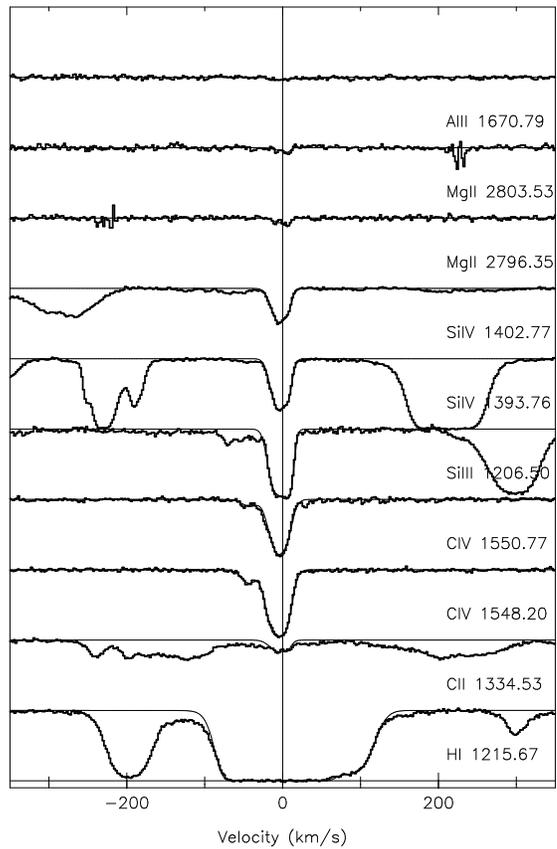}
\caption{Stacked velocity plot of heavy element absorption lines at
$z=2.168$.  See Section \ref{sec:z2.168} for description}
\label{fig:z2.16}
\end{figure}

Figures \ref{fig:z1.84} through \ref{fig:z2.57} show the raw data for
our absorption complexes, where we have shifted the various ions to
align vertically in redshift/velocity space.  When required for
clarity we have separated the low-ionization- and
high-ionization-potential species for individual systems into
different panels.

The zero point of the velocity scales are set to the redshift of the
nearest identified galaxy if one is known; when no galaxy is
identified we center on the strongest absorption component.  The thin
solid lines superimposed upon the data represent model Voigt profile
fits, made using the VPFIT
\footnote{http://www.ast.cam.ac.uk/$\sim$rfc/vpfit.html} software
package.  The output parameters are summarized in Table 2.

\subsection{Association of Lines According to Ionization Potential}\label{sec:ions}

All six systems exhibit transitions which vary widely in ionization
potential, but overlap in velocity space.  The most difficult aspect
of the fitting procedure was to distinguish which of these ions were
physically related, and which were aligned in velocity purely by
chance.  Intuition provided some guide, in that highly ionized \ovi or
\nv cannot arise in the same gas as nearly neutral species such as
\siii, \cii, or \feii.  This prior is supported by the velocity
profiles: the \ovi lines tend to be broad ($b\gtrsim 15$ km/s) and
smooth, whereas the low-ionization lines tend to be narrow ($b\lesssim
8$ km/s) and discrete.

The association of \civ and \siiv is less clear.  These ions bridge
the gap between the high and low potential species, sharing some
absorption components with each phase.  However, they also contain
components not associated with any other ion.  
In these cases, we looked to our
ionization models (described in Section \ref{sec:analysis}) for
guidance, iterating between the line fits and models to come upon a
physically plausible solution.  Our guidelines may be broadly
summarized with the following rules:
\begin{itemize}
\item{\ovi and \nv components are physically distinct from all lines
with lower ionization potential than \civ, including \siiv.  Although
\civ could coexist with \ovi, in practice we either do not see \civ at
the same redshift as \ovi, or the \civ has a different velocity
profile and is much too strong to have come from the same gas.}
\item{\feii cannot come from the same components as \siiv, \civ, or
higher ionization potenial lines.}
\item{\mgii and \alii can coexist with \siiv or \civ, but only in
trace quantities.}
\item{It is possible to observe \siii, \siiii, and \siiv, or \cii,
\ciii, and \civ simultaneously.}
\end{itemize}
Our tabulated absorption line fits all meet these criteria, but the
reader should be aware that they do not represent deterministically
unique solutions.  We simply established this short set of rules to
resolve the degeneracies in our data, with the requirement that the
results provide physically realistic inputs for our ionization models.

\subsection{Fit Technique, Comments on Robustness}\label{sec:fitting}

\begin{figure}
\plotone{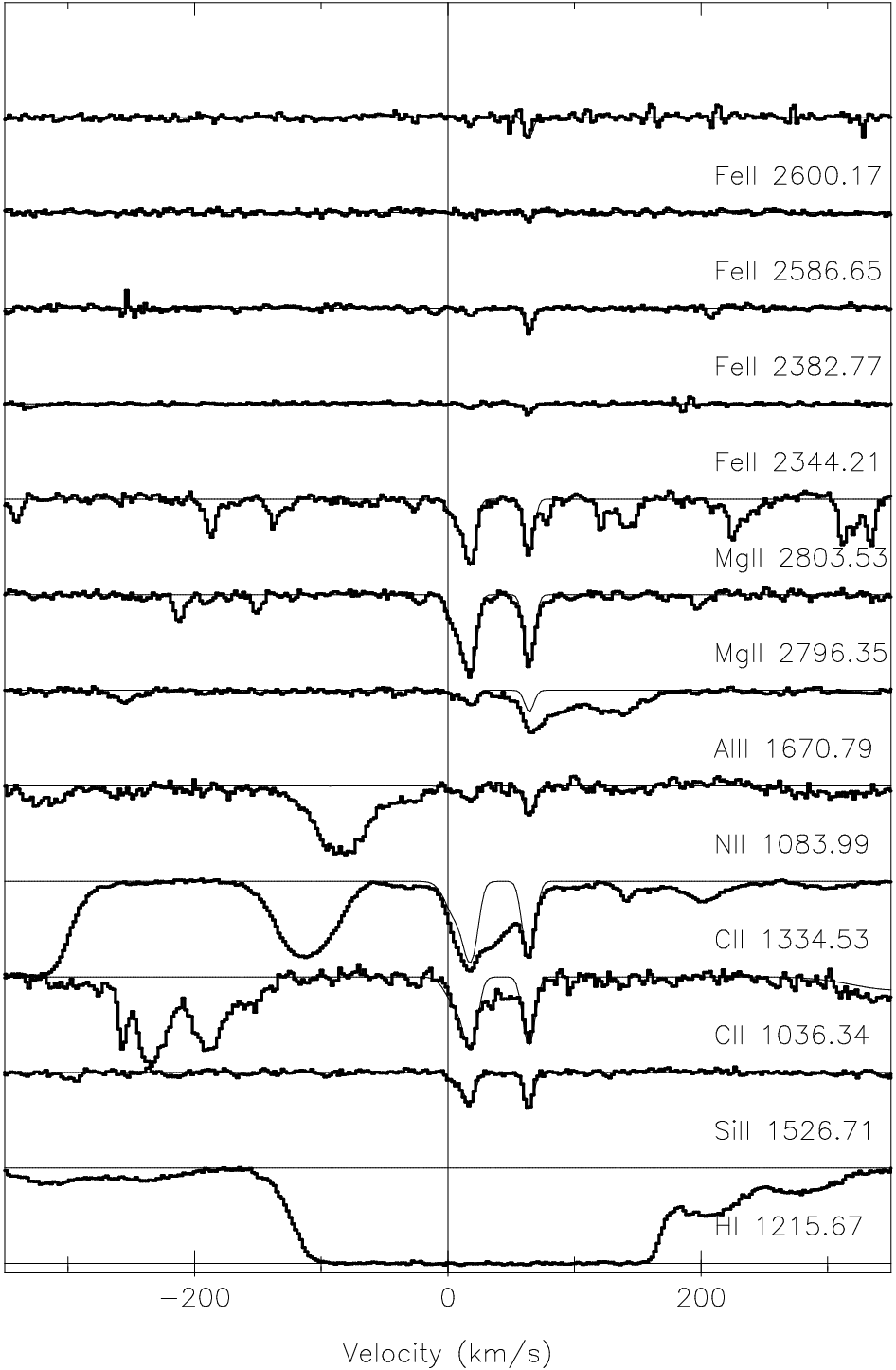}
\plotone{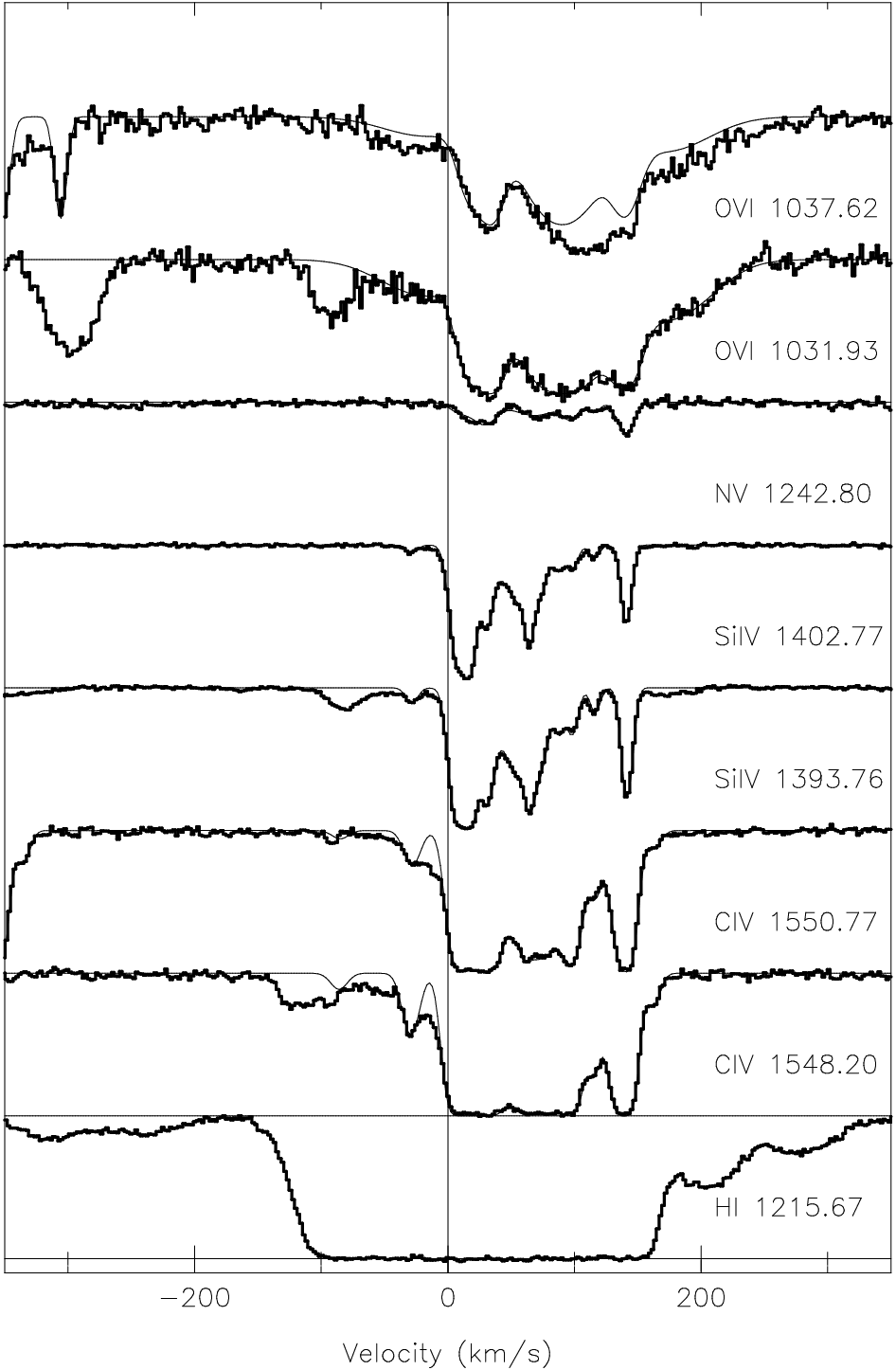}
\caption{Top: Low ionization lines for the system at $z=2.315$.  Zero
on the velocity scale corresponds to the redshift of the galaxy
Q1700-MD103, which has an impact parameter of $115\hinv$ kpc.  Bottom:
Highly ionized lines for the same system.}
\label{fig:z2.31}
\end{figure}

Our procedure has been to fit the singly ionized transitions first
(i.e. \cii, \siii, \mgii), since these typically have a small number
of narrow and easily discernable components.  When different ions
appeared to originate from the same gas phase (e.g. \cii and \siii),
we used a feature in VPFIT to ``tie'' the redshifts of their
individual components.  This enforces $z_\mcii = z_\msiii$ for the
tied components, but allows this redshift to vary during optimization.
This is reflected in the tables; when no redshift is listed for a
line, it is tied to its predecessor in the list.  It is also possible
to tie the $b$ parameters of fit components.  We used this approach
for ions from the same element (e.g. \cii, \ciii, and \civ).

For ions from different elements (with atomic masses $m_1$ and $m_2$),
the ratio of $b$ parameters can take on values ranging from $b_1/b_2 =
\sqrt{m_2/m_1}$ for purely thermal broadening, to $b_1/b_2\sim 1$, for
bulk or tubulent line broadening.  By measuring $b$ parameters of
elements with different masses, one can in principle solve for the
relative contributions of thermal and non-thermal motions to the total
line width.  This calculation is particularly interesting in our case
since both heating and turbulence in the IGM may reveal energy
dissipation from nearby starbursts.

Unfortunately, the $b$ parameters are not constrained well enough to
permit this type of analysis.  Where possible, we did attempt to fit
$b$ independently for different elements.  However, we encountered two
obstacles in interpreting the results.  First, the most metal-rich
systems had such narrow lines that they were under-resolved or
marginally resolved even by HIRES (resolution $\Delta v \approx 6.6$
km/s FWHM, corresponding to $b\approx 4$ km/s).  VPFIT accounts for
instrumental broadening by convolving a kernel with the model during
opimization, so in theory one can estimate linewidths for
under-resolved features.  However, we have modeled the line-spread
function only as a simple Gaussian.  This limits our confidence in the
measured linewidths for components with $b\lesssim 5$ km/s.  However,
since these low-ionization lines are very narrow to begin with, they
must surely be very quiescent: their temperatures must be very low
($T\lesssim 15,000$K), {\em and} their internal turbulent velocities
must be very small ($\Delta v_{\rm turb}\lesssim2-3$ km/s).

We also found it difficult to estimate $b$ for the higher ionization
\civ and \siiv systems because of line blending, and in some cases
saturation.  As noted previously, the profiles for these ions are very
complex, and they seem to trace a range of physical conditions.  In
the presence of strong blending, the Voigt profile models can take on
a range of solutions by trading column densities and $b$ parameters
between closely spaced components.  Typical uncertainties from these
degeneracies amounted to $\sim 0.1$ dex in column density and $2-4$
km/s in $b$.


The column density measurements for these systems are therefore much
more robust than the $b$ parameters.  For this reason we have only
quoted conservative upper limits on the temperature and turbulent $b$
parameters for most of the systems in Table 2.  However, there are a
small number of systems where our $b$ estimates are more secure, so we
list these values accordingly.

For highly blended systems, the quoted \civ and \siiv column densities
are derived assuming thermal relative line widths (indicated in table
footnotes).  We also fit the same systems with a purely non-thermal
ratio of $b$ parameters to gauge the effect on the column densities.
The fit quality ($\chi^2$) was similar for the two methods, and the
difference in column density was usually smaller than $0.1$ dex.

Our most difficult measurements were for \hi.  The \lya transition is
heavily saturated for all of the systems we are considering, yet all
are optically thin, as determined by FOS observations of the
high-order Lyman continuum \citep{HS1700_FOS}.  For several of the
higher redshift systems we could observe \lyb and other \hi
transitions, and in these examples the high order profiles contained
unsaturated regions.  However, the lines were usually still blended
and/or partially saturated, and at best we could estimate crude upper
limits on $\nhi$ from regions where the profiles (whose redshifts and
linewidths were inferred from metal lines) leaned up against
unsaturated pixels.  

Some of \hi column densities we report are smaller than the values
listed in the \citet{HS1700_FOS} FOS work; this is an important
discrepancy since these measurements are crucial for determining
absorber metallicities.  In the FOS data, $\nhi$ was determined by
measuring the discontinuity at the Lyman limit, at much lower
resolution ($\Delta v\sim 200-270$ \kms~ FWHM, compared with our $6.6$
\kms).  To compare our measurements with the the FOS data, we use the
{\em total} (i.e. summed) $\nhi$ for all of the high-resolution
subcomponents we have fit for each system.  For example, in the system
at $z=2.439$ we find a maximum component column density of
$\log\nhi=15.510$, whereas the FOS spectum yields $\log\nhi=15.890$.
However, at high resolution the system separates into six
subcomponents, whose aggregate $\log\nhi=15.870$ is in excellent
agreement with the FOS results.

The match is not always this close, but the median offset for the six
measured systems was $0.02$ dex, and in five of the six cases our
total $\nhi$ limit differed from the FOS value by less than 0.25 dex.
The one significant difference was for the system at $z=2.167$, where
our original $\nhi$ was $0.93$ dex lower than the FOS measurement.
For our HIRES measurement, we had assumed $b=35$ \kms, appropriate for
purely thermal line broadening relative to the heavy element lines.
However, Vogel and Reimers' curve-of-growth analysis yielded a
somewhat smaller value of $b=25$ \kms, and we also see some evidence
for turbulent line broadening in the relative heavy element
linewidths.  A choice of $25$ \kms~ does not conflict with the HIRES
profiles, and yields $\nhi$ values similar to the FOS results.
Accordingly, we have adopted these values as upper limits on $\nhi$
below.  In all other systems we quote the pure best-fit values from
the HIRES fits.  Although our \hi limits are quite rudimentary, they
are still internally consistent and provide important constraints on
the parameters of our ionization models---particularly the absorber
metallicities.


\section{Ionization Modeling of the IGM}\label{sec:analysis}

We have modeled the physical properties of our absorption line sample
using CLOUDY ionization simulations.  Since many of the complexes
contain absorption from a variety of elements and ionization states,
we can gain substantial leverage over the ion balance in individual
absorption components.


\begin{figure}
\plotone{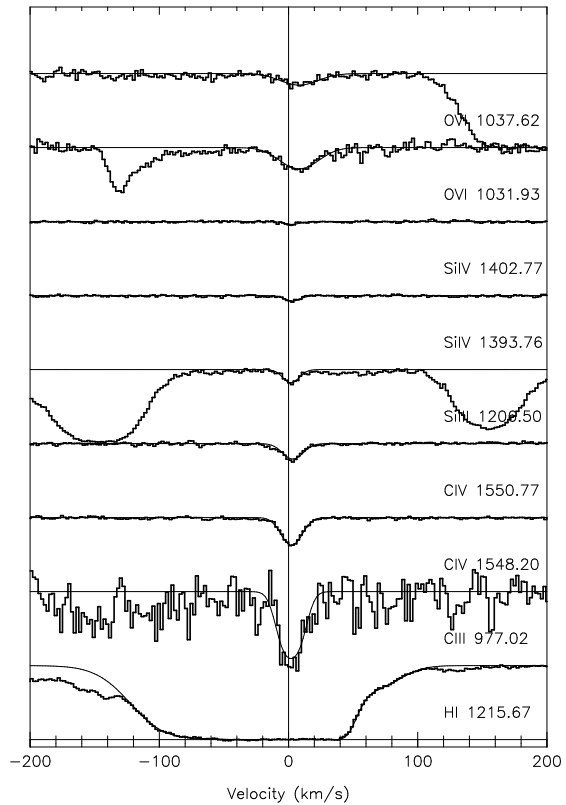}
\caption{Stacked velocity plot of heavy element absorption lines at $z=2.379$.  See Section \ref{sec:z2.379} for description.}
\label{fig:z2.37}
\end{figure}

\subsection{Radiation Field}\label{sec:localradiation}

We treated the absorbers as plane-parallel gas slabs, illuminated by
the integrated background light from quasars and galaxies.  The shape
of the background spectrum is derived from an updated calculation
based on the work of \citet{HM96}, and provided to us by F. Haardt.
The new spectrum uses a slightly different power law slope for the
shape of intrinsic QSO spectra ($f(\nu)\propto \nu^{-1.8}$), and it
also includes an integrated galaxy background.  The galaxies
contribute primarily at rest-frame optical and near-UV wavelengths,
but they also boost the flux blueward of the Lyman limit, where $10\%$
of photons are assumed to escape.  We used a separate background
spectrum for each absorber in the sample, with a shape and
normalization appropriate to its redshift \citep{scott_uv_bkgd}.

\begin{figure}
\plotone{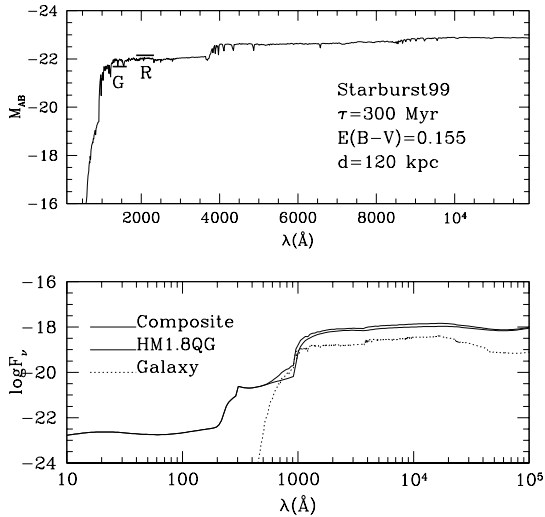}
\caption{Starburst background for MD103 ($z=2.315$).  To account for
the effects of local radiation, we added the spectrum of a starburst
galaxy to the standard Haardt \& Madau spectrum for the extragalactic
background field.  The template spectrum was a $300$Myr-old galaxy
taken from the Starburst99 archive, adjusted for dust reddening
according to a \citet{calzetti} law with $E(B-V)=0.155$.  This age and
reddening were chosen to match observations of $z\sim 3$ Lyman break
galaxies.  The spectrum was scaled to reproduce the observed $G$ and
$R$ band luminosities (Top panel), and then projected to a distance
corresponding to the galaxy's impact parameter from the QSO sightline.
The bottom panel shows the individual spectra of the galaxy (at the
location of the absorber), and the Haardt \& Madau background, and
finally the combined spectrum, which was used for all ionization
models of this system.  Most of the UV photons from the galaxy are
extinguished by dust, though there may be some local flux contribution
near the Lyman limit.}
\label{fig:MD103_bkgd}
\end{figure}

Since the effects of local ionizing sources may be important
\citep{miralda_local, schaye_local_ionizing}, we included a crude
local radiation model for the two absorption systems which are closest
to known high-redshift galaxies.  We downloaded synthetic spectra of
starburst galaxies from the Starburst99 archive \citep{starburst99},
with properties corresponding to constant star formation models with
an age of 300 Myr.  We included the effects of in-situ dust extinction
by applying the correction of \cite{calzetti}, with $E(B-V)=0.155$.
The corresponding escape fraction of Lyman limit photons is $\sim
30\%$, dropping off sharply to the blue.  The starburst age and
reddening parameters were chosen to match population models of
color-selected galaxies at $z\sim 3$ \citep{alice_lbg_nirc}.  The
synthetic spectra were normalized in absolute magnitude to match the
observed galaxy apparent magnitudes.  Then, the galaxy flux at the
location of the absorbing slab was estimated by applying a distance
modulus appropriate to the line-of-sight impact parameter for each
galaxy (Figure \ref{fig:MD103_bkgd}).  For the nearest system to the
sightline, ionizing photons from the galaxy outnumber those from the
metagalactic field by a factor of $\approx 2$; for all other systems
the radiation field at 912\AA ~was dominated by the diffuse
background.  One side of the slab was illuminated with this excess
local radiation.

\subsection{Methodology}\label{sec:CLOUDY_method}

Using these background spectra, we generated grids of CLOUDY models
whose parameters were varied to reproduce the column densities of each
individual absorption line.  For every model, we varied the total gas
density $n_H$, which in conjunction with the radiation field
determines a system's ionization balance.  We also varied the overall
metal content of the gas relative to the solar level, which we denote
as [X/H], and the relative [Si/C] abundance ratio.  The model results
are summarized in Table 3.  In the following discussion, we use the
term ``ion'' to denote a single absorption line from a single element
and ionization state.  We refer to a ``component'' as a {\em set} of
physically related ions from different elements, whose redshifts match
precisely.

\begin{figure}
\plotone{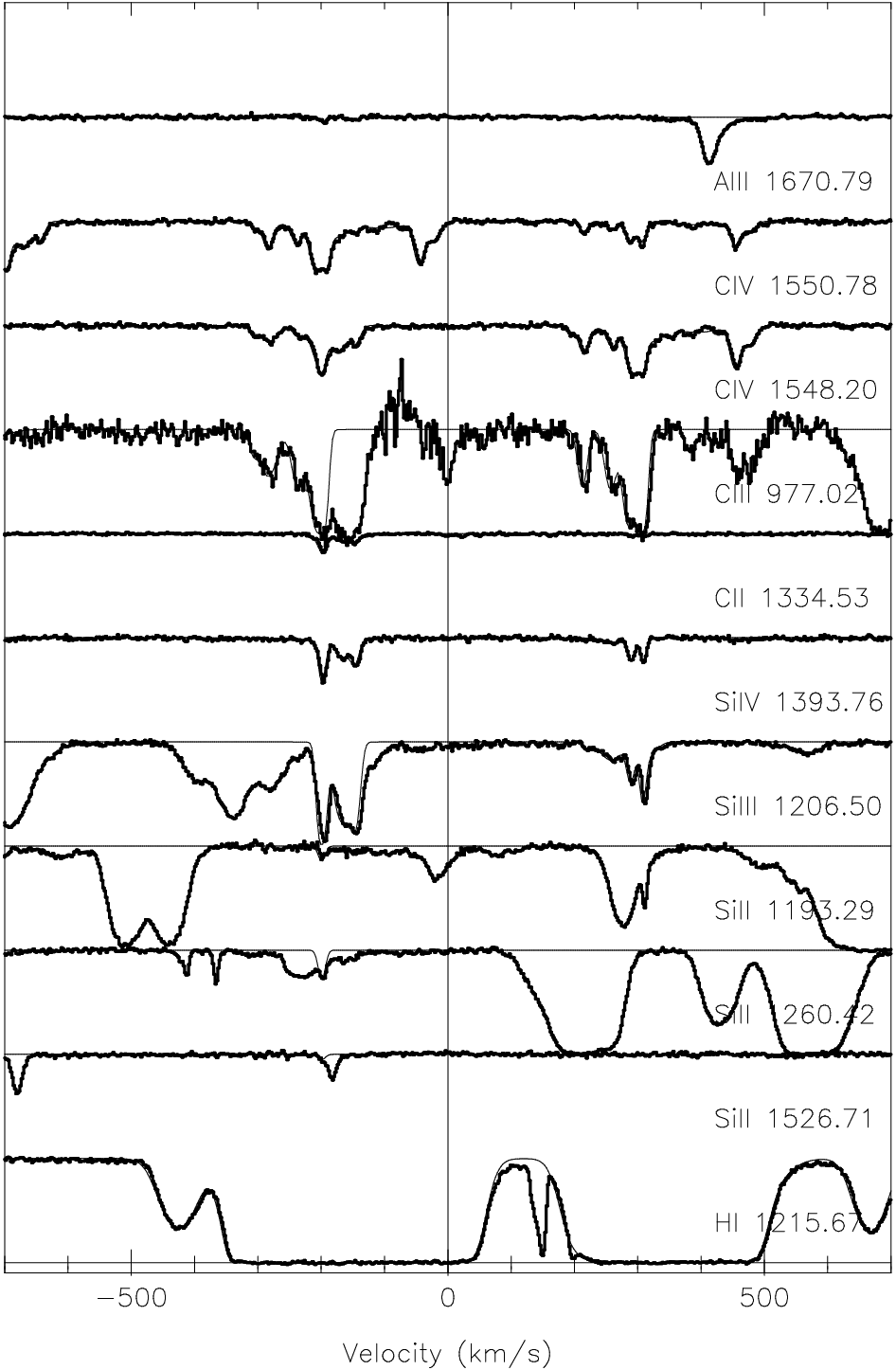}
\plotone{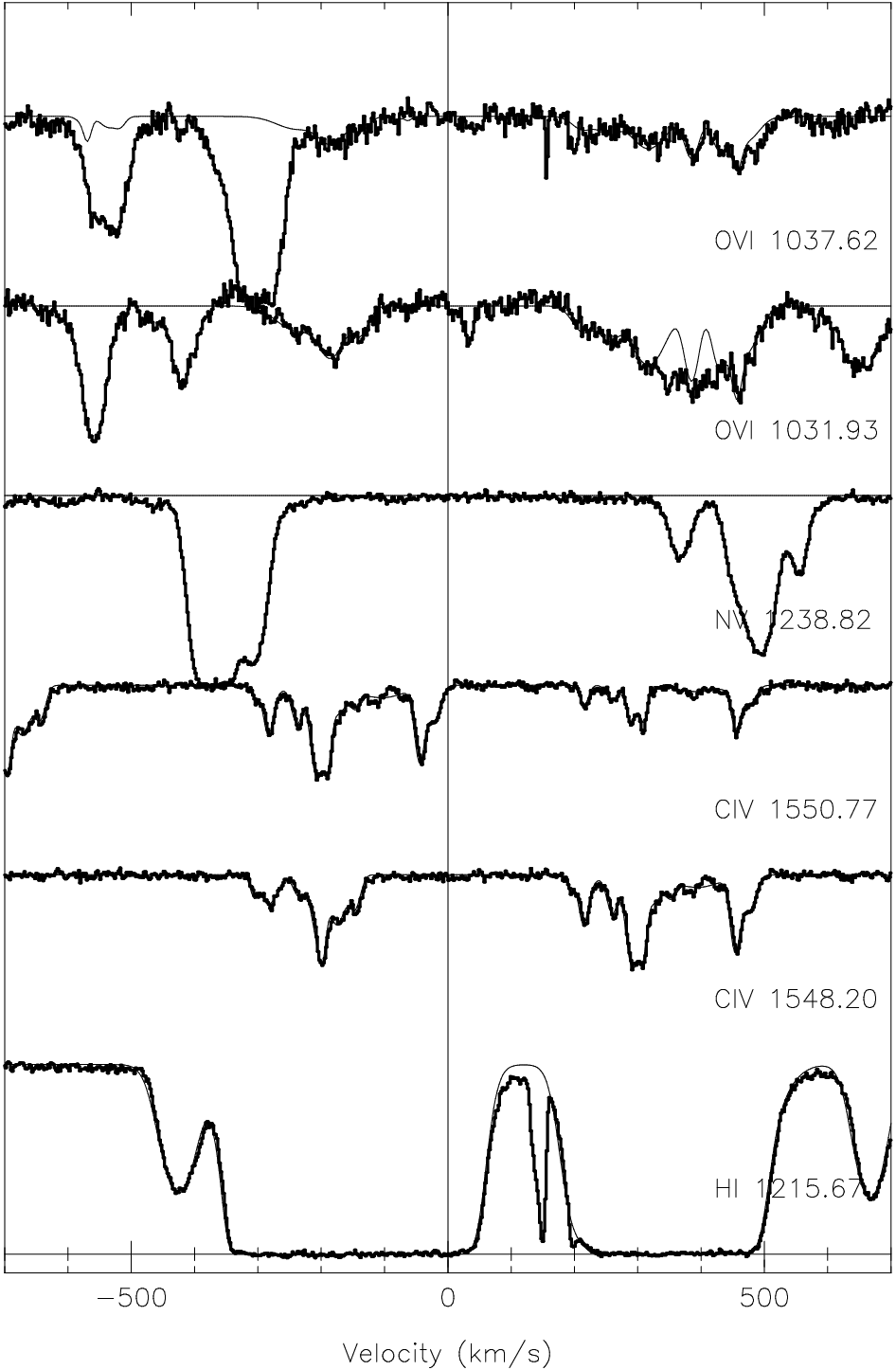}
\caption{Top: Low ionization lines for the system at $z=2.43$.  Zero
on the velocity scale cooresponds to the redshift of the galaxy
Q1700-BX717, which has an impact parameter of 210 kpc.  Bottom: Highly
ionized lines for the same system.}
\label{fig:z2.43}
\end{figure}

\begin{figure}
\epsscale{1.1}
\plotone{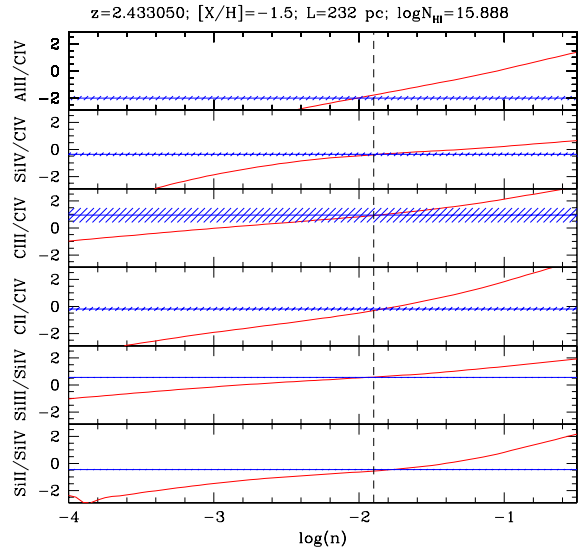}
\caption{Sample ionization model for a component in the complex at
$z\sim 2.43$.  Measured ion ratios with $1\sigma$ errors are shown by
blue horizontal lines with hatched regions.  CLOUDY predictions are
shown with red solid lines.  Vertical dashed line indicates the
density which yields the best match between the data and model.  This
system has $[X/H]=-1.5$ ($\sim \frac{1}{30}Z_\sun$), $n_H=10^{-1.9}$
cm$^{-3}$ ($\oden \sim 1000$), $\Delta L=232$ pc, and [Si/C]=$+0.37$}
\label{fig:cloudy_model}
\end{figure}

Figure \ref{fig:cloudy_model} demonstrates our methodology.  For each
component, we defined a set of observed column density ratios to be
used as model constraints.  Where possible, we favored ratios of ions
from the same element (e.g \cii/\civ, \siiii/\siiv) to eliminate
uncertainies arising from the relative abundances between different
elements.  We compared the selected ratios with their predicted values
from CLOUDY to determine the gas density.  For a given component, all
ion ratios typically matched near a common value of $n_H$, with only a
very weak dependence on metallicity.  This best-fit density
(determined via simple $\chi^2$) is indicated in the figure with a
vertical dashed line.

Next, we combined $n_H$ with CLOUDY-derived ionization fractions to
determine the absorbing pathlength for each heavy element ion in the
component:
\begin{equation}\label{eqn:pathlength}
L_{\rm ion} = {{N_{\rm ion}}\over{n_H (\frac{X}{H}) f_{\rm ion}}}.
\end{equation}
Here, $N_{\rm ion}$ represents the ion's (measured) column density,
$(X/H)=(X/H)_\odot\times 10^{[X/H]}$ represents the element's
abundance by number relative to hydrogen, and $f_{\rm ion}$ represents
the fraction of the element's atoms in the ionization state of
interest.  For a fixed radiation field, $f_{\rm ion}$ depends only
upon $n_H$ (i.e. $n_H$ determines the ``ionization parameter'').
After determining $L_{\rm ion}$ for each ion in a component, we
averaged the results together to obtain a mean pathlength
$\left<{L_{\rm ion}}\right>$ for that component.  Only heavy element
lines were included in the average since our \hi column densities are
poorly constrained.

We then inverted Equation \ref{eqn:pathlength} to produce model column
densities for each ion using the mean pathlength and best-fit model
parameters: $N_{\rm model}=n_H(\frac{X}{H})f_{\rm ion} \left<{L_{\rm
ion}}\right>$.  The scatter between these model column densities and
the observed values is our metric for model accuracy.  We minimized
$\chi^2$ between the model and data to determine optimal
metallicities, while requiring that the implied $\nhi$ satisfied
whatever upper limits on $\nhi$ were available.

For some systems, we were able to produce more accurate models by
varying the relative [Si/C] abundance.  This is particularly
interesting because [Si/C] enhancement is thought to be a signature of
enrichment from Type II supernovae.  We approached this measurement
cautiously, since a [Si/C] enhancement can be degenerate with a change
in gas density (ionization parameter) and we did not want to stretch
our model past what is warranted from the available data.  For this
reason, we only tested [Si/C] variations for components where we could
break this degeneracy by constraining the ionization parameter {\em
independently} using ratios from the same element, as in \siii/\siiv,
\siiii/\siiv, or \cii/\civ.  With the density and hence ionization
balances determined, we examined a single ratio ---\siiv/\civ---to
estimate [Si/C].

\subsection{CLOUDY Results}\label{sec:CLOUDY_results}

Table 3 summarizes the results from our ionization models.  For each
component, the best-fit values for the three model
variables---metallicity, density, and (if applicable) [Si/C]---are
listed first.  Then we show several cloud parameters that follow from
the optimal model, including $\left<{L_{\rm ion}}\right>$, $\nhi$, and
the equilibrium gas temperature $T_{\rm eq}$.  In the rightmost column
we list the the suite of ions used to constrain each component's
ionization model.

We have also quoted crude confidence intervals on the model parameters
for each component.  It is not straightforward to estimate
conventional errors for these model parameters, since ambiguities in
the Voigt profile fitting or model assumptions can compound in ways
that are not necessarily obvious.  However, we tested a wide range of
model parameters for each component\footnote{The full range of
parameters we examined was $-3\le[{\rm X/H}]\le0.5$, $-5.5\le n_H \le
-1.0$, and $-1 \le [{\rm Si/C}] \le 1$.}, and the data clearly
excluded significant portions of this search space in each case.  The
sub/superscript numbers in the table span the full range of values
permitted by the data.  They are not standard deviations propagated in
traditional fashion, but they do provide a feel for the model
uncertainties.

\begin{figure}
\plotone{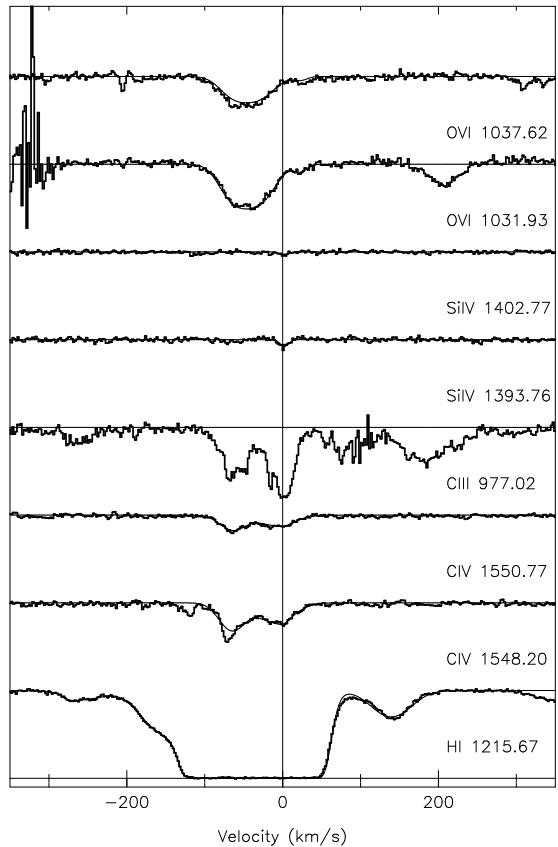}
\caption{Stacked velocity plot of heavy element absorption lines at
$z=2.578$.  See Section \ref{sec:z2.578} for description.}
\label{fig:z2.57}
\end{figure}

\subsection{Descriptions of Individual Galaxy/Absorber Systems}\label{sec:individual_systems}

In this section we provide a brief interpretation of the model output
for each system in the absorption sample, together with a description
of nearby galaxies.  We quote redshift differences in velocity units,
which may be mapped to distances using the Hubble parameter, which at
$z=2.3$ takes a value of $H=240\hinv$ km s$^{-1}$ Mpc$^{-1}$, not
accounting for peculiar velocities.

A key question concerns how far in impact parameter and/or velocity
separation we can allow galaxies and absorbers to be separated while
still considering them to be physically associated.  Theoretical
models of supernova-driven winds at $z\lesssim 5$ tend to find
stalling radii near $\sim 100$ kpc, with the limit set by a
combination of energetics and travel time
\citep{aguirre_z3_winds,fujita_feedback_simulations,
bruscoli_feedback, juna_feedback, benson_feedback}.  Especially given
the large mass estimates for the galaies in our sample \citep[$M_{\rm
halo}\sim 10^{11-12}M_\odot$,][]{adelberger_clustering}, it may be
difficult to eject much interstellar mass to larger distances.
However, outflowing material is seen directly in high redshift galaxy
spectra, blueshifted by $\Delta v\sim 350\pm250$ \kms~ relative to the
stellar rest frame.  While it is possible that the winds contain hot
gas moving at higher speeds (up to $\sim 1000$ \kms), the material
with the largest geometric cross-section for QSO absorption should
also be moving tangential to the QSO sightline, so its projected
line-of-sight velocity should be systematically smaller at larger
impact parameters.  In the following discussion, we consider strong
galaxy/absorber associations to be systems where the galaxy and
absorber are separated by $\lesssim 200$ kpc in impact parameter
and/or $\Delta v \lesssim 600$ \kms in velocity.  A more detailed
justification of these choices will be presented further below in
Section \ref{sec:absorbers_near_galaxies}.

\subsubsection{$z=1.846$ (Figure \ref{fig:z1.84})}\label{sec:z1.846}
 
This complex was identified via \nv, but it also contains a number of
other, lower ionization potential lines.  We do not identify any
galaxies with this absorber; the closest candidate (BX635), is still
quite far at $327\hinv$ proper kpc from the quasar sightline and
$1500$ km/s in redshift space.  The \cii, \siii, and \mgii absorption
is narrow and unresolved---this system is very cold and kinematically
quiescent.  It was difficult to measure $\nhi$ since our spectrum only
covered \lya.  We estimated upper limits by fixing \hi components at
the redshifts of the metal lines, setting their $b$ parameters to the
values appropriate for thermal broadening, and increasing $\nhi$ to
match the wings of the absorption.  The resulting $\nhi$ matched the
FOS measurements of \citet{HS1700_FOS} to within 0.1 dex.  The FOS
spectrum shows some evidence of \ovi absorption, though the doublet
appears to suffer from blending at the observed resolution.  Our
best-fit CLOUDY models find a near-solar metallicity, with a minimum
plausible value of $\sim \frac{1}{3}Z_\sun$.  The gas density is near
$10^{-2.5}$ cm$^{-3}$, or $\oden\sim 300$ times the cosmic mean, and
the absorbing pathlengths are of order $\Delta L\sim 100$ pc.  The
dense environment and high abundances suggest that this system was
enriched very recently by an unidentified galaxy that is either
fainter than $R=25.5$, or whose colors fell outside of the $UGR$
selection space (See Section \ref{sec:galaxy_obs}).

\subsubsection{$z=2.168$ (Figure \ref{fig:z2.16})}\label{sec:z2.168}

This system was identified based on the detection of several carbon
and silicon transitions, along with \mgii and \alii.  It is ambiguous
whether this system contains \ovi, since the entire \ovi region is
saturated by interloping \lyb lines.  However, there is clearly no
\nv.  There is also no obvious galaxy association; the nearest
candidate is BX691 which has an impact parameter of $\rho=290\hinv$
kpc but $\Delta v=+2000$ km/s from the absorber.  Our initial fits to
the HIRES data (assuming thermal broadining of \hi relative to the
metal lines) found $\nhi\lesssim 10^{15.9}$ for $b=35$ \kms.  For
these values the CLOUDY model finds a metallicity of $\sim
\frac{1}{10}$ solar or higher, with $\Delta L\sim 100$ pc and
$\oden\sim 300$.

However, as described in Section \ref{sec:fitting}, measurements of
the Lyman limit discontinuity using FOS imply $\nhi\sim 10^{16.8}$ and
$b=25$ \kms.  The carbon and silicon linewidths in the HIRES spectrum
show evidence of turbulent line broadening at the $b_{\rm turb}\sim
10-15$ \kms ~level; this would be sufficient to bring the total \hi
$b$ parameter and column density into a range consistent with FOS.
Adoption of the higher $\nhi$ value changes the model metallicites
substantially, allowing values as low as [X/H]$\sim-2.5$, at sizes of
$1$ kpc and larger, and densities of $\oden\sim 100-1000$.  Given the
sensitivity of $\nhi$ to the assumed $b$ parameter, the metallicity
for this system is fairly uncertain.  However, it does appear to be
quite over-dense.  It may be mildly enriched by a galaxy that remains
unidentified, or it may represent a particularly dense and stirred-up
gas cloud with a metallicity comparable to the surrounding IGM.

\subsubsection{$z=2.315$ (Figure \ref{fig:z2.31})}\label{sec:z2.315}

This is the strongest system in the absorption sample, and it is
located at the same redshift as the closest galaxy to the QSO
sightline.  MD103 is one of the more luminous galaxies near HS1700.
Stellar population models incorporating rest-frame IR photometry from
Spitzer/IRAC find a stellar mass of $M_*\sim 10^{11}M_\sun$, and a
(somewhat uncertain) star formation age of $t_{\rm sf}\sim 300-1000$
Myr \citep{alice_spitzer}.  The stellar mass ranks MD103 among the
top $\sim 10-15\%$ of galaxies at $z\sim 2.3$.  It also has a large \ha
equivalent width, implying a substantial star formation rate---in the
range of $64-88 M_\odot$ yr$^{-1}$ \citep[dust
corrected,][]{erb_halpha}.

Strong \ovi absorption reveals hot gas in the IGM surrounding MD103.
We also detect \nv, though it appears to be kinematically associated
with \civ rather than \ovi.
Mixed in with this material are small regions of very cool, very
dense, very metal-rich material.  These components are narrow and
unresolved, but they contain absorption from \feii, which is only seen
in extremely dense environments.  Their ionization models yield solar
metallicity, $\oden\sim 10,000$, and sizes of $\sim 1$ pc.  There are
also many components seen in \civ and \siiv, which seem to trace gas
with intermediate temperatures and densities.  This combination of
shock-heating, vigorous cooling, high densities, solar metal
enrichment, and a star-forming galaxy, all suggest that the HS1700
sightline is piercing a feedback mixing zone where debris from MD103
is being deposited into the IGM at $R\sim 100$ kpc.

\subsubsection{$z=2.379$ (Figure \ref{fig:z2.37})}\label{sec:z2.379}

This is the simplest system in our absorption line sample.  None of
the observed galaxies can plausibly be linked with the IGM absorption.
Our best-fit CLOUDY model predicts a metallicity of $\sim
\frac{1}{100}$ solar, densities near $\oden\sim100$, and sizes of
several kpc for the component traced by carbon and silicon.  There is
also a clear \ovi detection with no corresponding \nv.  The highly
ionized component is slightly offset in redshift, and can be explained
equally well by a hot, collisionally ionized plasma, or a low density
($\oden \lesssim 10$), large ($\Delta L\sim 10-100$ kpc) photoionized
structure.  This complex probably represents a clump of material
embedded in a modestly enriched intergalactic filament.  Its
metallicity is only $\sim 1\sigma$ above the cosmic median, and there
is little else to suggest vigorous feedback in the immediate
neighborhood.

\subsubsection{$z=2.43$ (Figure \ref{fig:z2.43})}\label{sec:z2.43}
 
This absorption system is coincident in redshift with the galaxy
BX717.  Like MD103, BX717's luminosty is at or slightly below average
for $z\sim 2.5$ galaxies, but it is forming stars at a rate of $\sim
20M_\odot$ yr$^{-1}$ \citep{erb_halpha}.  Population models yield an
integrated stellar mass of $M_*\sim 10^{10}M_\odot$ for BX717, with a
stellar age of $400-500$ Myr \citep{alice_spitzer}.  The QSO
absorption system has a distinctive double-troughed profile, with the
strongest components situated approximately $\Delta v=-200$ km/s and
$+350$ km/s from BX717.  It is tempting to interpret this as a
signature of expanding walls of a wind bubble originating from the
galaxy.  However, given BX717's somewhat large impact parameter
($\rho=216\hinv$ physical kpc) it is at least as likely that the
double troughed shape is merely coincidental.

The CLOUDY models yield intermediate metallicities near
$\sim\frac{1}{30}$ solar---about a factor of 10 higher than the IGM.
The gas densities are in the $\oden \sim 1000-3000$ range, with
absorbing pathlengths of several hundred parsecs to a kiloparsec.
We detect broad, but somewhat weak \ovi absorption with no
correponding \nv, which we interpret as hot gas with $T\sim 300,000$
K.  
It is difficult to say whether the heavy elements originated in BX717,
or in an unidentified companion galaxy.

\subsubsection{$z=2.578$ (Figure \ref{fig:z2.57})}\label{sec:z2.578}

This simple system has an unusually broad and featureless \civ
profile, and a strong \ovi line.
The best fit model yields a pathlength of several kpc, with
overdensity $\oden\sim 50-100$ and clear [Si/C] enhancement.  The \ovi
absorption is most likely produced in hot gas, though a low density,
photoionized solution is possible if the structure is extremely large
($L\gtrsim 100$ kpc).  The nearest galaxy (MD119), is offset in
redshift by $\Delta v \sim 1000$ km/s and has a fairly large impact
parameter ($\rho=336\hinv$ physical kpc).  MD119 is probably not
responsible for the QSO absorption, though it may be embedded in a
related larger structure.  In fact this system resembles an unusually
metal-rich filament or group environment more than a shock-heated
wind.

\subsection{Overview of Model Results and Comments on Robustness}\label{sec:overview}

\begin{figure}
\epsscale{1.1}
\plottwo{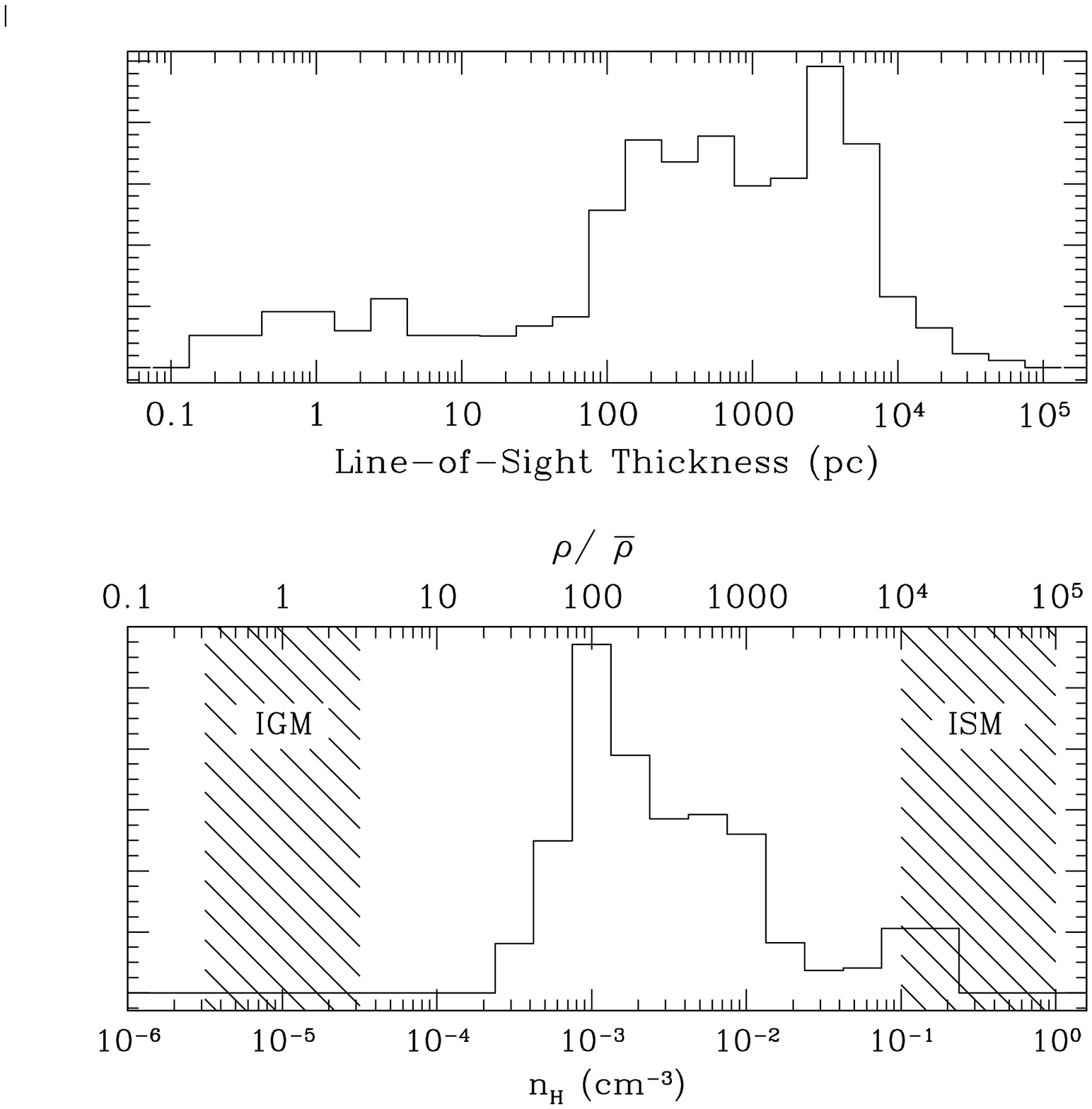}{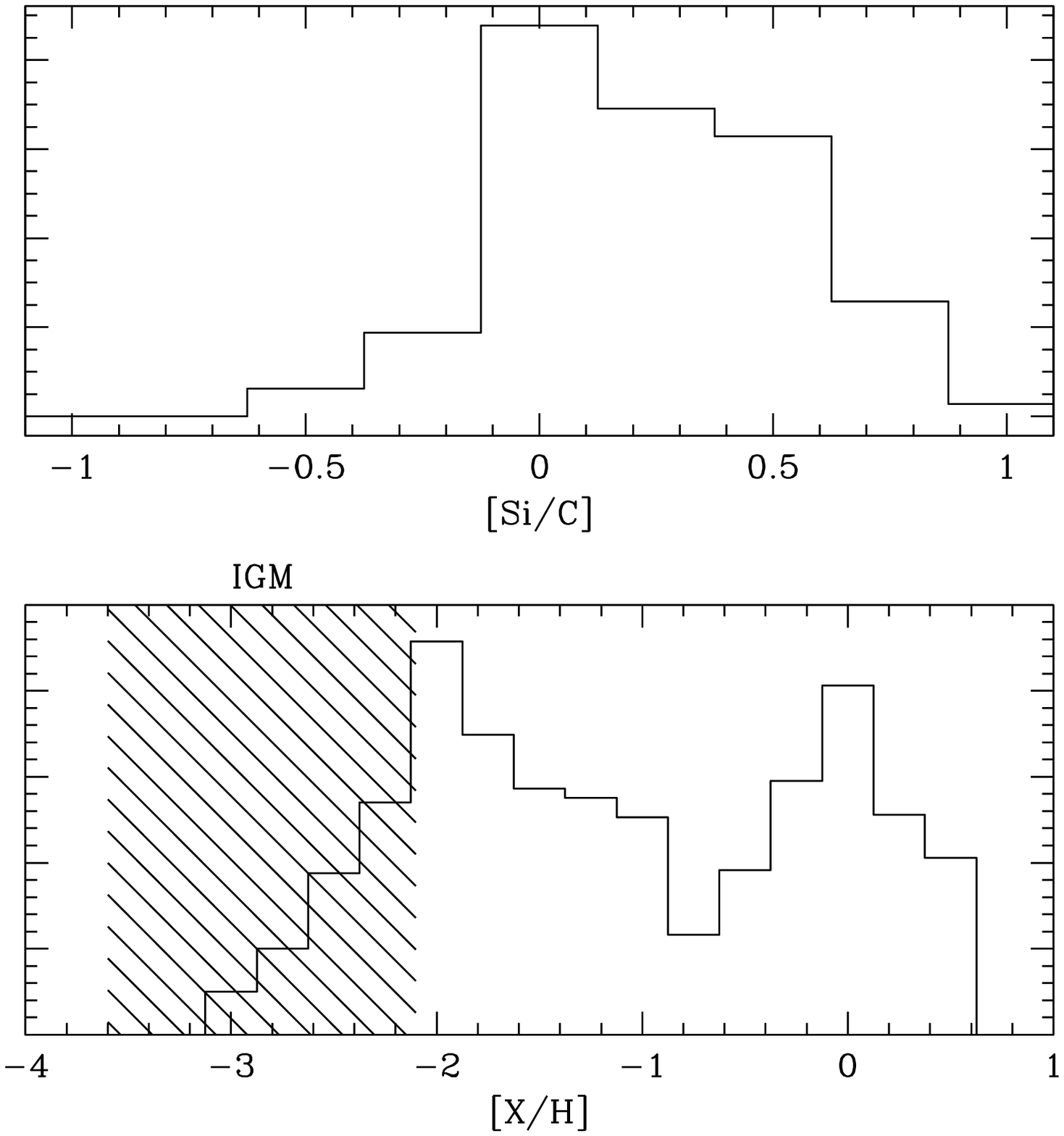}
\caption{Parameter histograms for our CLOUDY models of absorption
components.  These are not simple histograms of the best-fit values;
rather, for each component we account for the contribution throughout
the uncertainty interval quoted in Table 3, assuming that the
probability distribution is flat within this range.  Our sample
selects absorbers with small thicknesses ($L\sim 1$ kpc), high
densities ($\oden \gtrsim 100$), and [Si/C] abundance enhancements.
The metallicities are distributed well above the intergalactic median,
and show evidence of bimodality.}
\label{fig:cloudy_hist}
\end{figure}

Figure \ref{fig:cloudy_hist} summarizes the distribution of model
parameters for our full absorption line sample.  We have constructed
these histograms using a slightly unusual method.  Rather than
assigning each component to a single bin, we spread its contribution
over the full range of bins encompassed by its errors, normalizing so
that its total addition to all bins equals one unit.  Thus, a
component with large model uncertainties adds a small amount to many
bins, while a well-constrained system may add a single count to a
single bin.  The final distributions approximate probability densities
for the parameters, marginalized from our multi-dimensional search
space.

The most noticable result in Figure \ref{fig:cloudy_hist} is a
tendency toward small cloud sizes and high metallicities.  For the
majority of our systems the best-fit metallicity ranges from $0.01-1
~Z_\sun$, with some evidence for bimodality in the distribution.  For
comparison, the general IGM has a median abundance of $\sim
\frac{1}{700}Z_\sun$; its $\pm 1\sigma$ abundance contours for $z\sim
2.5$ are shaded in the figure \citep{simcoe2004,schaye_civ_pixels}.
It appears that these absorption systems are chemically polluted
relative to the universe at large.  Systems in the upper half of the
bimodal distribution have abundances of at least $\frac{1}{3}Z_\sun$.
This makes them even more metal rich than the interstellar gas
contained within many high redshift galaxies
\citep{lbg_winds,cb58_pettini,alice_K20_metallicity}.  They are most
likely the product of recent galaxy formation and feedback.  The lower
half of the metallicity distribution may trace slightly older galactic
debris which has been diluted upon mixing into the IGM, or even true
intergalactic gas which was pre-enriched by much earlier episodes of
star formation.

Line-of-sight absorber thicknesses of a few hundred parsecs to a few
kpc are typical, though some of the most metal-rich systems are even
sub-parsec in scale.  These sizes are consistent with the {\em
transverse} scales of \civ systems derived from multi-sightline
analysis of lensed QSOs \citep{rauch_civ_lens}. The absorbing
structures must therefore be quite small, or contain substantial
substructure if they are organized into larger units.  The best-fit
gas densities are several orders of magnitude higher than the mean
density of the IGM at $z\sim 2.5$, but somewhat lower than the
characteristic density of galactic interstellar media.


\begin{figure}
\epsscale{1.1}
\plotone{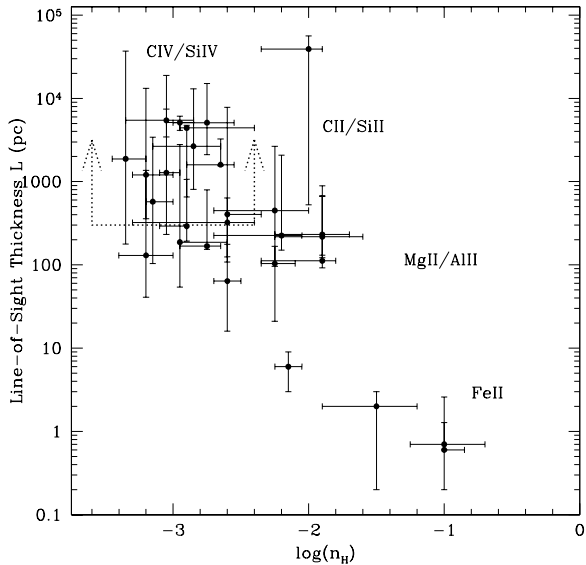}
\caption{Correlation between absorber density and ``size''
(line-of-sight absorbing pathlength).  
There is a rough sequence of absorption line properties following from
upper left to lower right in the diagram.  The largest, lowest density
absorbers are highly ionized and seen exclusively in \civ and/or
\siiv.  Moving downward, one observes species with increasingly lower
ionization potential, initially picking up \cii and \siii, then \mgii
and \alii, and finally \feii in the smallest, densest pockets.  Dotted
arrows indicate the smalllest approximate transverse sizes of \civ
absorbers, estimated from double-sightline observations of lensed
quasars \citep{rauch_civ_lens}.}
\label{fig:density_size}
\end{figure}

In the systems where we could measure [Si/C], there is some evidence
for a relative silicon abundance enhancement by 0.1-0.5 dex.  This is
roughly consistent with heavy element yield calculations for Type II
supernovae, which predict a similar [Si/C] boost
\citep{woosley_weaver_yields,umeda_nomoto_yields,chieffi_limongi_yields},
so it is possible that the metal-rich absorbers are preferentially
enriched by debris from the explosion of high-mass stars.

\begin{figure}
\plotone{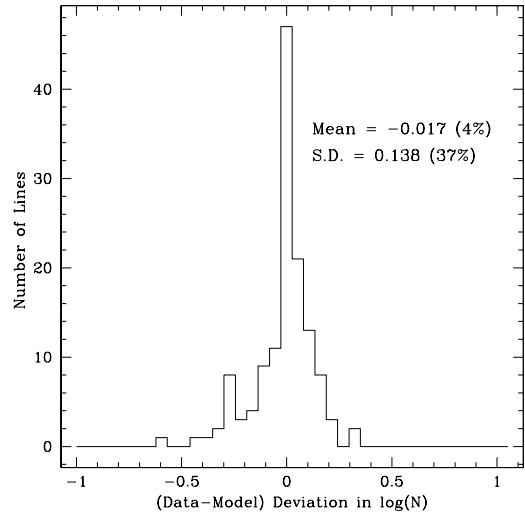}
\caption{Histogram of the difference between $\log (N_{\rm meas})$---the
Voigt profile column density for each ion in the sample---and the
best-fit model predictions from CLOUDY.  The models on average
reproduce the data with no statistically significant systematic
offset.  Expressed in linear terms, the typical scatter for an
individual line is $\sim 37\%$.}
\label{fig:cloudy_errors}
\end{figure}

Though we have displayed the model results as histograms, in practice
the parameters can be correlated.  For example, in Figure
\ref{fig:density_size} we show a scatter plot of $n_H$ versus $\Delta
L$ which illustrates how the properties of the metal-rich absorbers
fall along a rough sequence.  At the top of this sequence are the \civ
and \siiv components, which have sizes of one to several kpc and low
densities.  Moving to higher densities one encounters smaller
structures traced by lower ionization potential lines, from
\cii/\siii, to \mgii/\alii, and finally \feii at extremely small sizes
and high densities.  This sequence is a useful tool for estimating the
physical conditions in a given absorption system at a glance, based on
what low-ionization lines are visible.  The absence of points in the
lower left corner of the plot is a selection effect; we would not be
sensitive to very small, low density systems.  However, systems in the
upper right quadrant should be detectable if they are present.  A
larger sample containing Lyman limit and damped \lya systems might
populate this corner of the plot.  Their absence in this small sample
illustrates that the optically thin, metal-rich absorbers are more
common than their optically thick counterparts, and hence must have a
larger statistical cross-section.

Among the model parameters we have measured, the gas density $n_H$ is
most robust, since it is determined solely from ratios of heavy
element ions and mostly insensitive to variations in metallicity in
the optically thin limit.  The uncertainty in metallicity was larger,
since its primary constraint comes from difficult \hi measurements.
For a small number of systems the heavy element ratios changed just
enough with metallicity to rule out very low values.  However, we
usually used the upper limit on $\nhi$, measured from unsaturated
regions of the \lyb or higher order profiles, to determine a lower
bound on [X/H].  Even with such weak \hi constraints, the
corresponding lower limits on [X/H] are often still significant.

A measure of the models' accuracy is the overall scatter between their
predicted ion column densities and the true observations.  A histogram
of this metric is shown in Figure \ref{fig:cloudy_errors}, and the
individual model column densities are listed at the far right of Table
2 for comparison with the true measurements.  We see no statistically
significant systematic offset between our model column densities and
the data, but there is a residual RMS scatter of $\sim 0.14$ dex.
This reflects the combined error from the Voigt profile fits, the
model assumptions, and from within CLOUDY itself.  Put another way,
our models reproduce the observed column densities with a {\em linear}
accuracy of $\sim 37\%$ or better.

\section{Physical Properties of the Metal-Rich Absorption-Line Systems}\label{sec:physical_model}

So far we have shown that our absorption sample---identified primarily
by strong \ovi, but supplemented with one \nv and one \mgii
system---identifies compact, dense clouds with unusually high heavy
element abundances.  In the next two sections we consider a simple
explanation for these properties, involving the propagation of
bubble-like shocks through intergalactic filaments.  The absorbers'
high abundances strongly suggest that the shocks are being expelled
from---rather than falling into---star forming regions.

\subsection{Arguments for a Sheet- or Shell-Like Absorber Geometry}\label{sec:shells}

To see even a handful of metal-rich absorption systems in a single
sightline, the aggregate cross-section of the population must be
significant.  We estimate their average cross-section in the standard
fashion:
\begin{equation}
N=n\sigma{{c}\over{H_0}}\Delta X
\end{equation}
where $N$ is the number of detections in the quasar spectrum, $n$ is
the comoving number density of the absorbing structures, and $\Delta
X$ is the absorption pathlength probed along the
sightline\footnote{The pathlength is given by \\ $\Delta X =
(1+z)\left[{\Omega_M(1+z)+\Omega_\Lambda/(1+z)^2}\right]^{-\frac{1}{2}}\Delta
z$ for $\Omega = 1$.  We searched the range $1.8\le z\le 2.7$ for a
total $\Delta x =2.9$.  The search boundaries were determined by
signal-to-noise ratio considerations in the blue, and a desire to
avoid the proximity effect near the QSO in the red.}.  Assuming a
circular cross section, $N=3$,
\footnote{For the following analysis we exclude the systems at
$z=2.379$ and $z=2.568$, since they have lower heavy element
abundances and gas densities than the most metal-rich systems in the
sample.  We also exclude the system at $z=2.168$; its gas density is
high, but its metallicity is poorly constrained and could be
substantially below solar levels.} $\Delta X=2.9$ for our HS1700
spectrum, and comoving number density $n$, we calculate the
cross-sectional physical radius as:
\begin{equation}\label{eqn:shell_xsection}
R \sim 190 ~h_{71}^{-1}~{\rm kpc} \left({{{n}\over{0.0021 ~{\rm
Mpc^{-3}}}}}\right)^{-\frac{1}{2}}.~~~{\rm(Shell ~geometry)}
\end{equation}
Here we have normalized the comoving number density to that of $R\le
25.5$, UV-selected, $z\sim 2.3$ galaxies as determined by
\citet{adelberger_clustering}.  Clearly this cloud radius $R$ is much
larger then the inferred absorber thickness $\Delta L\sim 1$ kpc (see
Figure \ref{fig:cloudy_hist}).  Apparently if one associates the
metal-rich absorbers with luminous galaxies, their geometry must
resemble a thin sheet or shell.

Alternatively, if the absorption arises from a random distribution of
much smaller, spherically symmetric clouds with $\sigma \sim \pi
(\Delta L/2)^2$, the implied comoving density is extremely large:
\begin{equation}
{{n_{\rm abs}}\over{n_{\rm gal}}} \sim 150,000 \left({{\Delta L}\over{1 ~{\rm kpc}}}\right)^{-2} ~~~{\rm (Cloudlet ~geometry)}
\end{equation}
i.e., the absorbers outnumber known $z\sim 2.3$ galaxies by over 5
orders of magnitude.  The enhancement can be severe, since the
smallest systems in our sample have $\Delta L\sim 1$ pc but are seen
in many of the metal-rich complexes.  For comparison, the Local Group
contains $\sim 20-25$ known low mass objects per $L^*$ galaxy.
Another perspective comes from simple Press-Schechter (1974)
considerations \citep[e.g,][]{mo_white_2002}: if observable galaxies
at $z\sim 2.3$ have total masses of $M\sim 10^{12}M_\sun$
\citep{adelberger_clustering}, then the mass of the dark matter haloes
that outnumber these galaxies by a factor of $\sim 10^5$ is $\lesssim
10^6 M_\sun$---smaller than globular clusters.

These statistics are reminiscent of the ``weak'' \mgii absorbers seen
locally \citep{churchill_weakmgii, rigby_weakmgii}, and the arguments
for their sizes, metallicities, and number densities are essentially
identical.  In fact several of our systems would qualify as weak \mgii
systems; the complex at $z=2.315$ in particular resembles the
``iron-rich'' class of weak \mgii absorbers, which have $\sim 1$ pc
scale absorption depths and near-solar metallicities.  However, in the
high redshift data there is evidence of a connection between these
cool, dense systems and strong \ovi absorption.  This is an important
clue that the metal-rich systems were recently shock-heated and have
subsequently cooled into their present state.

Small dense cloudlets like these would not be in pressure equilibrium
with the IGM.  Their characteristic gas density of $\log(n_H)\gtrsim
-3$ represents an enhancement of $\oden\gtrsim 100$ relative to the
cosmic mean at $z\sim 2.3$, even as their optically thin interiors are
maintained at a similar temperature as the exterior IGM through
photoionization ($T\sim 5,000-15,000$ K for clouds versus $T\sim
20,000-30,000$ K for the IGM).  The cloudlet pressure therefore
exceeds that of the IGM by a factor of $\sim 10-100$.  This will cause
them to puff into the IGM on a timescale comparable to their sound
crossing time, which for a system with $R\sim 500$ pc and $c_s\sim 16$
km s$^{-1}$ is roughly 30 Myr.  Even if the clouds are embedded in
$10^6 M_\sun$ dark matter haloes, the gravitational potential is too
weak to confine the gas, since the sound speed ($\sim 16$ km s$^{-1}$)
exceeds the halo escape velocity ($\sim 5-10$ km s$^{-1}$) even well
within the halo.  It would be very difficult to self-enrich such a
halo to solar levels, since it is doubtful that the baryons would
survive even a single supernova event.

These arguments favor the sheet or bubble scenario over one involving
a large population of small cloudlets.  The shells have high chemical
abundances and are mixed with hot \ovi gas, suggesting shock heating
and a recent association with star-forming environments.  In other
words, the metals at large galactocentric radii probably originated in
the interstellar medium of nearby galaxies and were ejected violently
into the IGM.  The presence of luminous galaxies within $\sim
250\hinv$ kpc of half the absorbers bolsters this hypothesis.

Any realistic model of these sheets or bubbles would probably not be
symmetric or monolithic, since the shocks propagate through a
non-uniform medium and may develop hydrodynamic instabilities.
Clumping of material within the sheets would lead to somewhat larger
radii in Equation \ref{eqn:shell_xsection}.  However, this only
reinforces the basic result: sheets or bubbles provide the most
efficient geometry to create a large absorption cross-section from
structures with small linear dimensions.

\subsection{Radiative Shocks as a Likely Origin for Metal-Rich Absorbers}\label{sec:shocks}

A natural location to form thin shells or sheets is at the interface
of shock fronts, particularly when the shocks have become radiatively
efficient.  Order-of-magnitude calculations show that a metal rich,
radiative shock propagating through the IGM will generate conditions
much like those observed in the metal-rich absorbers.

During the early evolution of a strong shock when radiative losses are
insignificant, $\approx75\%$ of the shock's kinetic energy is used in
heating the post-shock material, so $T=\frac{3}{16}\frac{\mu m_H}{k}
v_{\rm shock}^2$, which with $m_H$ as the proton mass and $\mu=0.62$
(for fully ionized gas) yields:
\begin{equation}\label{eqn:shock_temperature}
T\approx 10^6 ~K \left({{v_{\rm shock}}\over{265 ~{\rm km ~s}^{-1}}}\right)^2.
\end{equation}

At $T\gtrsim 10^6$ K, the cooling timescale for post-shock gas can be
quite long.  However, in time the shock will decelerate (e.g. from ram
pressure, or from gravity and $pdV$ work in the case of galactic
winds) and the post-shock temperature will descend into the peak
region of the cooling curve, near 300,000 K.  This transition occurs
at $v_{\rm shock}\sim 145$ km s$^{-1}$.  At this point, the shock
becomes radiatively efficient, and a thin dense shell is formed which
travels at the bulk speed of the shock front.  In a radiative
(isothermal) shock, the pre- and post-shock temperatures are identical
since the shock's thermal energy is radiated away on timescales that
are short compared to the system's evolutionary timescale.

In this situation, the post-shock gas density relates to the pre-shock
value as $n_{\rm post}\approx \mathcal{M}^2n_{\rm pre}$, where
$\mathcal{M}=v_{\rm shock}/c_s$ is the Mach number, i.e. the shock
front velocity in units of the sound speed in the undisturbed medium.
By continuity across the shock boundary, we also have $v_{\rm
post}\approx v_{\rm pre}/\mathcal{M}^2$.  In the $z\sim 2.5$ IGM, the
sound speed is roughly $c_s\sim \sqrt{kT/\mu m_H}\sim 16$ km s$^{-1}
T_{20}^{0.5}$, where $T_{20}$ is the temperature of the general IGM
normalized to 20,000 K \citep{schaye_thermal}.  Since the shock becomes radiative at $v\sim
145$ km/s, the Mach number at the time of this transition is
$\mathcal{M}\sim 9 T_{20}^{-0.5}$, which yields the following values
for the density and post-shock velocities:
\begin{eqnarray}\label{eqn:post_shock_params}
n_{\rm shell} & \sim & 10^{-2.1} ~{\rm cm}^{-3}\times \nonumber \\
 & &\left({{n_{\rm IGM}}\over{10^{-4} ~{\rm cm}^{-3}}}\right)\left({{T_{\rm IGM}\over{20,000 {\rm K}}}}\right)^{-1} 
\left({{\mathcal{M}}\over{9}}\right)^2\\
v_{\rm post} & \sim &  1.8 ~{\rm km ~s}^{-1} \left({{T_{\rm IGM}\over{20,000 {\rm K}}}}\right)
\left({{\mathcal{M}}\over{9}}\right)^{-2}.
\end{eqnarray}
Here we have assumed that the shock is travelling through an
intergalactic filament with $\oden=10$ relative to the cosmic mean
density $\bar{n}\approx 10^{-5}$ (for $z\sim 2.5$).  Note that $v_{\rm
post}$ is not the speed of the shock, but rather the speed of
post-shock gas in the frame of the shock, which is still moving at
over 100 \kms.  The number density $n_{\rm shock}$ for filaments with
$\oden\sim 1-10$ may be compared with the observed distribution of
$n_H$ shown in Figure \ref{fig:cloudy_hist}.  For consistency we can
also verify that the cooling timescale of the shocked, ionized gas is
short:
\begin{eqnarray}
\tau_{\rm cool}&\sim& {{3n_{\rm shell}kT}\over{n_{\rm shell}^2\Lambda}} \\
&\sim& 0.5 ~{\rm Myr} 
\left({{n_{\rm IGM}}\over{10^{-4} {\rm ~cm^{-3}}}}\right)^{-1}
\left({{T_{\rm IGM}}\over{20,000 {\rm ~K}}}\right)^{-1} \nonumber \\
& & \times \left({{v_{\rm shock}}\over{145 {\rm ~km~s^{-1}}}}\right)^{2}
\left({{\Lambda(T)}\over{10^{-21} {\rm ~erg~cm^{3}~s^{-1}}}}\right)^{-1}
\end{eqnarray}
where $\Lambda(T)$ is the gas cooling function from
\citet{sutherland_dopita} evaluated at 300,000 K (or the temperature
given by Equation \ref{eqn:shock_temperature}).  This rapid
thermalization with the background radiation field lends {\em ex post
facto} support to the CLOUDY models developed in Section
\ref{sec:analysis} under the assumption of photoionization
equilibrium.

A crude estimate of the shell thickness is given by the product of the
shock's propegation timescale and post-shock gas velocity.  If the
shocks are produced during galaxy assembly, and have timescales
similar to the star formation ages of $z\sim 2.5$ galaxies
\citep{alice_spitzer}, we find:
\begin{eqnarray}
\Delta L&\sim& v_{\rm post} \tau_{\rm sf} \\
&\sim & 550 ~{\rm pc}
\left({{T_{\rm IGM}}\over{20,000 {\rm ~K}}}\right)
\left({{\tau_{\rm sf}}\over{300 {\rm ~Myr}}}\right)
\left({{\mathcal{M}}\over{9}}\right)^{-2}
\end{eqnarray}
This distance is also comparable to the line-of-sight absorber
thicknesses shown in Figure \ref{fig:cloudy_hist}.  Evidently, the
large compressions achieved in radiative shocks give rise to compact,
high-density shells with densities and thicknesses similar to the
observed values.

Finally, using the above calculations we may estimate the \hi column
density of the radiative shock front:
\begin{eqnarray}
N_\mhi &\sim& n_{\rm shell} \cdot \Delta L \cdot f_\mhi\\
       &\sim& 10^{16.5}~{\rm cm}^{-2} \nonumber \\
& & \left({{\tau_{\rm sf}}\over{300 {\rm ~Myr}}}\right)
{\left({{n_{\rm IGM}}\over{10^{-4} {\rm ~cm^{-3}}}}\right)}
\left({{f_\mhi}\over{10^{-2.5}}}\right)
\end{eqnarray}
Here we have made use of our CLOUDY grid to determine the proper value
of $f_\mhi$ for the post-shock density given in Equation
\ref{eqn:post_shock_params}, and solar heavy element abundances.  This
column density is generally consistent with our observations, though
it is on the high end of the distribution.  This discrepancy is easily
reconciled if the gas metallicity is reduced even to $Z/Z_\sun\sim
0.1$ (thereby lowering $f_\mhi$), or if the ambient filament
overdensity is reduced below $\oden \lesssim 10$.  The important point
is that these systems have much higher column densities than typical
\lya forest lines, but they are still optically thin in \hi.


The strong shocks present at earlier times should produce hot material
that has a long cooling time.  The \ovi absorption in our sample may
be a residue from this earlier stage.  In principle one could also
catch a young system before its shock begins to radiate, in which case
it would be visible in \ovi but not in \hi or lower ionization lines.
We did not detect this type of system in a survey of several quasar
sightlines \citep{simcoe2002}, but they may exist in smaller numbers.
The relative rarity of \ovi-only QSO absorbers is naturally explained
by galactic wind models since the shell's cross section grows as it
cools.  This is particularly true if the radiatively-inefficient shock
phase is short lived.

The mere presence of \ovi absorption does not require one to interpret
a particular absorption system as a feedback candidate.  In fact we do
not report \ovi detections for two systems in our sample, though its
presence is not ruled out in either case (one exhibits \nv but has a
redshift too low for ground-based \ovi measurements, while the other
is blanketed by interloping \lyb absorption).  One can alternatively
produce \ovi via accretion shock heating as in the low redshift IGM
\citep{dave_whim}, or through pure photoionization of low density gas
\citep[e.g.][]{chaffee86}.  In fact, we have argued that some of the
systems in our sample arise from some of these very mechanisms.
However, the combination of (1) strong but optically thin \hi, (2)
strong absorption from both high and low ionization potential carbon,
silicon, magnesium, and iron, (3) highly ionized gas, and (4) an
association with galaxies together provides a strong case for the
feedback interpretation in several of the systems in our sample.

\section{Discussion}\label{sec:discussion}

We have presented data and ionization models for six strong absorption
systems in the HS1700+6416 sightline.  Five of these were selected
based upon the detection of highly ionized gas (\ovi and/or \nv); the
sixth was identified by its weak \mgii absorption.  Half of
these systems are very metal-rich, with abundances ranging from
$\frac{1}{30}$ solar to solar levels.  Each of the metal-rich
absorbers contains small ($\lesssim 100$ pc) pockets of very dense,
enriched material.  We have also shown that these systems' number
statistics and absorption properties are consistent with an origin in
thin, radiatively efficient shocks.  This, together with their high
abundances, suggests that these systems represent 
debris from galaxy formation, where material ejected from the
proto-galaxy's intersteallar medium is mixing into the nearby IGM.

\subsection{The Galaxy Environment of High-Metallicity Absorbers}\label{sec:galaxies_near_absorbers}

Luminous galaxies are located near over half of the metal-rich systems
(2 of 3), at essentially identical redshift and small impact parameter
($\rho\lesssim 200\hinv$ kpc).  These galaxy-absorber systems are also
reported by \citet{adelberger_z2}, based upon the same dataset.  The
strongest system ($z=2.315$, MD103) appears to be associated with the
closest galaxy to the quasar sightline, though it is still more than
100 kpc distant.  If the metal-rich systems result from winds or
merger feedback, then roughly equal numbers of feedback zones
(weighted by cross-section) at $z\sim 2.5-3.5$ may have come from
luminous Lyman-break-type objects, and unidentified galaxies which are
either fainter than $R=25.5$ or have different colors than known
$z\sim 2.5$ galaxy populations.

Several investigators have uncovered evidence for correlations between
luminous galaxies and \civ absorbers both in the local universe
\citep{chen_civ_galaxies} and afar \citep{kurt_winds, adelberger_z2}.
This absorption has variously been ascribed either to accretion of
pre-enriched intergalactic gas \citep{haehnelt_civ,rauch_simulations}
or dwarf satellites \citep{chen_civ_galaxies} during hierarchical
galaxy assembly, or to the expulsion of metal-rich material from
star-forming galaxies \citep{kurt_winds,adelberger_z2}.

It is somewhat unlikely that the present absorption sample is produced
by accretion alone; the observed metal abundances ($Z/Z_\sun\sim
0.1-1$) are higher than one would expect in such a scenario
($Z/Z_\sun\sim 10^{-2.8}$).  Simulations of gas infall tend to
reproduce the properties of general \civ systems
\citep{haehnelt_civ,rauch_simulations}, but they do not always follow
the high-end tail of the metallicity distribution traced by our
sample.  We argue below that infall of material that is pre-enriched
at high redshift ($z \gtrsim 5$) may explain the majority of \civ
systems in the ``field,'' but that the strongest metal line
systems---particularly those containing \ovi---are probably related to
more recent galaxy formation.

If this interpretation is correct, then some of the enriched gas is
probably expelled from galaxies slightly below the magnitude limits of
present high redshift surveys (recall that half of the absorbers do
{\em not} have an associated galaxy).  While even bright nearby
galaxies could remain unidentified due to color effects, there are
reasons relating to wind energetics that a fainter population might be
favored.  Since less massive galaxies are more numerous, the
absorption cross section becomes smaller according to Equation
\ref{eqn:shell_xsection}; at the same time the smaller mass results in
lower halo escape velocities.  Dark matter haloes in the range $M_{\rm
DM}\sim 10^{10}-10^{11}M_\sun$ should outnumber known $z\sim 2.3$
galaxies by about an order of magnitude \citep{mo_white_2002}, which
would reduce the bubble radius to $R\sim 70$ kpc---a distance well
within reach for theoretical models of supernova-driven galactic winds
\citep{aguirre_z3_winds}.

Alternatively, to enrich the IGM to near-solar levels through purely
dynamical processeses, one would need to shred the ISM of a forming
galaxy during the merger/assembly process, but after several
generations of stars have had time to enrich the gas.  The
characteristic velocities \citep[several hundred \kms,
][]{barkana_infall} and timescales (several hundred Myr) should be
similar for superwinds and slingshot mergers.  However, the merger
would need to be arranged so as to distribute its debris fairly
uniformly, rather than in one-dimensional tails which have small
absorption cross-sections.  One possible solution to this problem
involves the adiabatic expansion of winds whose thermal energy is
provided by shock-heating from mergers, rather than supernovae
\citep{merger_feedback}.


The lower metallicity systems in our absorption sample have lower gas
densities and larger sizes ($\Delta L\gtrsim 5$ kpc line-of-sight).
These systems resemble moderately enriched large-scale structures, and
have metallicities within $1\sigma$ of the mean level observed in the
\lya forest.  Their \ovi absorption may arise in accretion shocks
associated with collapsing structures or they may trace quite large
($L\sim 100$ kpc) photoionized structures---our models cannot
distinguish between these two possibilities.  No galaxy candidates
have been identified for these systems.

\epsscale{0.9}
\begin{figure}
\plotone{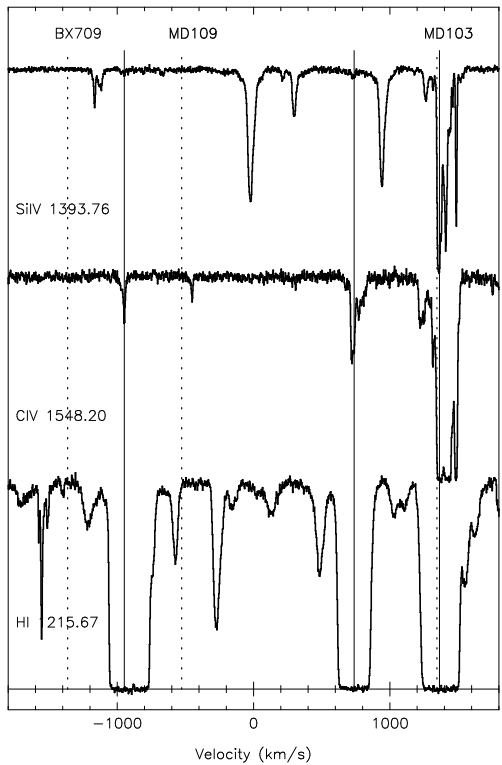}
\caption{\hi, \civ, and \siiv portions of HS1700 absorption line
spectrum, centered on the redshift of galaxy proto-cluster reported by
\citep{q1700_spike}.  The velocity range shown approximately matches
the redshift extent of the galaxy overdensity.  We detect three metal
absorption line systems in the cluster neighborhood, indicated with
solid vertical lines.  Redshifts of galaxies within $\rho\le
500h_{71}^{-1}$ physical kpc of the QSO sightline are indicated with
vertical dashed lines and labeled accordingly.  The systems at -100
and +800 km/s are typical of ``field'' \civ absorbers in terms of
\civ/\hi and \siiv/\civ ratios.  They do not have galaxies that
correspond closely in redshift, although the system at -1000 km/s is
flanked by two galaxies that may be embedded in a common large-scale
structure.  The system at +1400 km/s is clearly much more metal-rich,
and is characteristic of the sample of radiative shock systems
explored in this paper.  We do not detect diffuse \ovi or \nv that
would indicate shock-heated intra-cluster gas.}
\label{fig:q1700_spike}
\end{figure}

\subsection{A Galaxy Cluster Near an \ovi System}\label{sec:q1700_cluster}

There is evidence that one reported galaxy-absorber pair lies at the
outskirts of a proto-cluster at $z\sim 2.310$ \citep{q1700_spike}.
This raises the possibility that the shock-heated \ovi gas represents
intra-cluster matter rather than halo gas from the nearest individual
galaxy.  The few galaxy clusters known at $z\gtrsim 2$ do not resemble
the virialized, relaxed structures seen in the local universe, though
they may subsequently evolve into such a state
\citep{lbg_spike,pentericci_cluster,kurk_highz_clusters,venemans_cluster}.
The high redshift clusters exhibit clear galaxy overdensities, but
they are somewhat large both on the sky ($L\gtrsim 10$ Mpc) and in
redshift space ($\Delta z \sim 0.05, \Delta v \gtrsim 3000$ \kms).

In Figure \ref{fig:q1700_spike} we plot the absorption profiles of
several ions throughout the entire galaxy overdensity ($2.285\le z\le
2.315; \Delta v \sim 3000$ \kms).  We detect three significant \lya
absorbers with \civ at $\Delta v\sim -900, +750,$ and $+1400$ km
s$^{-1}$ relative to galaxy overdensity's center at $z=2.300$.  The
detection of several significant \lya absorbers is consistent with the
results of \citet{kurt_winds}, who also find large \lya opacities near
proto-clusters at $z\sim 3$.  However, it is clear that the system at
$\Delta v\sim +1400$ km s$^{-1}$ is qualitatively different from the
other two \lya absorbers; in fact this is the MD103 galaxy/absorber
complex described in Section \ref{sec:z2.315}, underscoring the
distinctive nature of the \ovi-selected metal line systems.

The other two systems are also strong \hi absorbers with $\nhi\sim
10^{16}$, but they have much lower \siiv/\civ ratios indicating a
higher degree of ionization, and lower \civ/\hi which indicates a
lower heavy element abundance ($[X/H]\lesssim -2.0$).  They are not
accompanied by hot \ovi or \nv.  The system at $\Delta v \sim -1000$
km s$^{-1}$ (relative to the cluster center) is flanked both on the
sky and in redshift space by a pair of galaxies (MD109 and BX709) with
impact parameters of $\rho=208$ and $316$ kpc and $\Delta v=\pm 400$
km s$^{-1}$ (relative to the absorber).  Their orientation on the sky
is indicated in Figure \ref{fig:q1700_field}.  This configuration
could easily result from the galaxies being embedded in a large
filament whose gas has been pre-enriched at earlier times to a level
near the intergalactic mean.

There are several weak \hi lines with $\nhi\sim 10^{12}$ cm$^{-2}$
dispersed throughout the galaxy overdensity, but we do not detect any
diffuse \ovi or \nv gas that would indicate the presence of a shocked
intra-cluster medium (ICM).  There are no absorbers at the central
redshift of the galaxy overdensity.  On the whole, the data do not
show evidence of uniform, highly enriched gas in the proto-cluster
environment at $z\sim 2.3$.  However, roughly solar abundances are
observed near the closest galaxy to the line of sight.  This suggests
that the metals and shock heating are powered by dynamical processes
associated with the galaxy itself rather than interactions with the
cluster.  If MD103 and other galaxies in the congealing proto-cluster
are losing their interstellar media through winds or dynamical
interactions, this could produce substantial intra-cluster enrichment
even before the ICM appears to have virialized.

\subsection{The Absorption Environment of Color-Selected Galaxies}\label{sec:absorbers_near_galaxies}

\begin{figure}
\epsscale{1.25}
\plotone{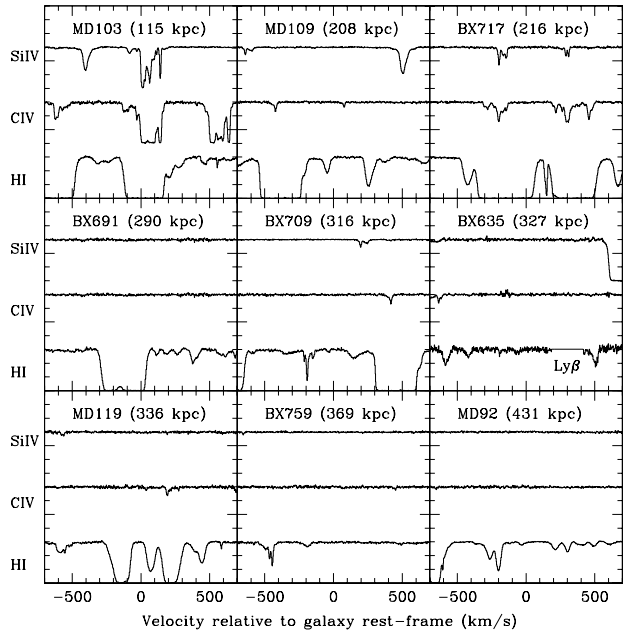}
\caption{Montage of \hi, \civ, and \siiv absorption in the vicinity of
 the $z\sim 2.5$ galaxes listed in Table 1.  The systems are arranged
 from top left to bottom right in order of increasing impact
 parameter.  Beyond $\sim 320h_{71}^{-1}$ kpc, the absorption
 properties resemble typical \lya forest regions with no distinct
 metal absorption. We have removed interloping \lya lines from the the
 \civ and \siiv profiles of BX635 (see text).}
\label{fig:galaxy_grid}
\end{figure}

Having described the galaxy population in absorption-selected
environments, we now consider the inverse problem: characterization of
the IGM near a sample of color-selected galaxies.  Figure
\ref{fig:galaxy_grid} shows a montage of absorption line plots
centered on the redshifts of nine galaxies from Table 1.  The plots
are sorted according to galaxy-absorber impact parameter, from upper
left to lower right.  For each panel we show the profiles of \hi,
\civ, and \siiv.

There appears to be a qualitative change in the properties of the IGM
at galaxy-absorber impact parameters $\rho \sim 320\hinv$ kpc.  Inside
this radius, all five detected galaxies have strongly saturated \lya
absorption at $\Delta v \lesssim 500$ \kms; four of the five are also
\civ systems.  At impact parameters larger than $\sim 320\hinv$ kpc,
we see no trends in the absorption data that would suggest an
association with galaxies.  The only absorption near BX635 is \lyb
from a higher redshift system (which we have masked in the figure).
The raw \civ and \siiv profiles from BX635 are contaminated by
interloping \lya; we verified that these lines could not be masking
true \civ or \siiv, since at least one of the doublet components was
free of \lya across the profiles.  For presentation purposes, we fit
the profiles of the interloping \lya near BX635 and removed their
signatures.  MD119 exhibits fairly unremakable \lya forest lines, and
BX759 and MD92 have very weak \hi profiles.  Except for MD119 the
systems at $\rho\ge 320\hinv$ kpc exhibit no heavy element absorption.

\begin{figure}
\epsscale{1.25}
\plotone{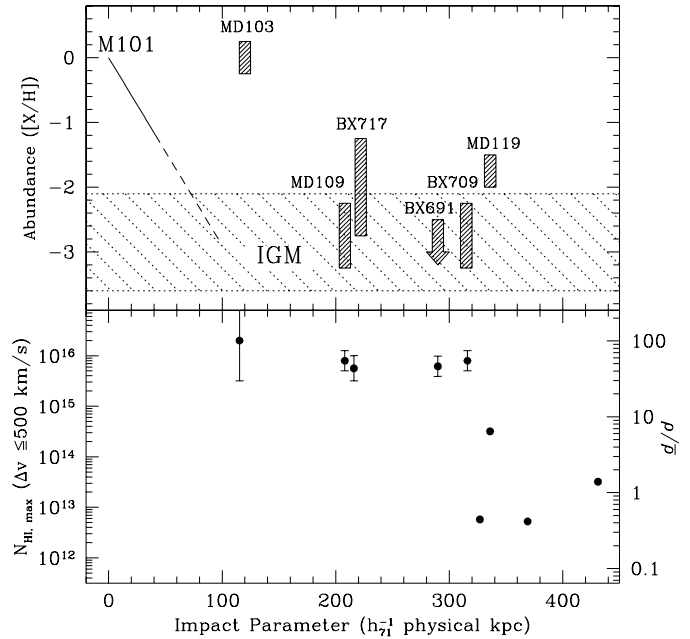}
\caption{Variation of gas density (bottom) and chemical abundance
(top) in the IGM as a function of galaxy impact parameter.  The gas
density shows evidence of a marked decline at impact parameters of
$\sim 300\hinv$ kpc, which we interpret as the transverse scale of gas
structures where the galaxies are embedded.  Errors in $\nhi$ for
points at large radii are smaller than symbols.  The chemical
abundances are very high at small impact parameter, but blend into the
background field (shown with hatched bar) between $100-200\hinv$ kpc.
The large intergalactic metallicity near MD103 cannot be explained by
extrapolation of in-situ disk enrichment trends---for reference, we
show the abundance gradient of the local galaxy M101's HII regions as
a function of galactocentric radius \citep{kennicutt_m101}.  Solid
line represents measurement area, while the dotted line is an
extrapolation of the measured trend.}
\label{fig:impact_metals}
\end{figure}

Figure \ref{fig:impact_metals} shows a graphical representation of
these trends.  In the bottom panel, we plot the column density of the
strongest \hi line located within $\Delta v = 500$ km/s of each galaxy
redshift, as a function of galaxy/absorber impact parameter.  There is
an apparent dropoff in \hi column at $\rho \sim 300\hinv$ kpc, with
smaller impact parameters exhibiting $\nhi\sim 10^{15.5-16}$ and
larger ones showing $\nhi\sim 10^{12.5-13.5}$.  Using the formulae of
\citet{schaye_forest} one can translate these \hi column densities
into approximate baryonic overdensities; this scaling is shown at the
right of the bottom panel.  Galaxies close to the sightline tend to
reside in regions with $\oden\sim 30-100$, whereas separations of
$\gtrsim 300\hinv$ kpc tend to yield densities near or slightly below
the cosmic mean.

The simplest interpretation is that the galaxies are embedded in
gas-rich intergalactic structures with transverse scales on the sky of
$\sim 300$ physical kpc ($\sim 1$ comoving Mpc), similar to the scales
reported by \citep{adelberger_z2}.  The structures are revealed
through moderately strong \lya forest absorption with either no heavy
element lines (e.g., BX691), or abundances consistent with the cosmic
mean (e.g., BX709, MD109).  In other words, luminous galaxies reside
in regions resembling randomly chosen, moderately overdense filaments,
as one would expect from hierarchical structure formation models.

The top panel of Figure \ref{fig:impact_metals} shows the variation of
heavy element abundance in the IGM with galaxy impact parameter.  The
metallicity estimates were made in the manner described in Section
\ref{sec:analysis}, using multiple ion measurements coupled with
CLOUDY models.  The data's sensitivity did not permit abundance
measurements in the low density IGM (i.e. beyond $300\hinv$ kpc), nor
could we measure abundances lower than $[X/H]\sim -3$.  However, the
systems that we measured show some interesting trends that might not
be expected from hierarchical structure formation alone.

In the $200\lesssim \rho\lesssim 320\hinv$ kpc range which is most
heavily populated by our sample, the chemical abundances are
comparable to general \lya forest systems, whose mean $\pm1\sigma$
contours are shown in the plot with a horizontal hatched bar
\citep{simcoe2004, schaye_civ_pixels}.  Though individual systems
appear to scatter above the mean, we remind the reader that we cannot
measure the low-metallicity end of the distribution (e.g., we only
measure an upper limit for BX691).

The one galaxy in our sample with $\rho \lesssim 200\hinv$ kpc (MD103)
exhibits a marked enhancement in metallicity above the general IGM.
This $\sim 300$-fold abundance jump is not consistent with a simple
extrapolation of in-situ galactic disk enrichment trends.  For
comparison we show the radial abundance gradient of M101, one of the
best-studied disk galaxies in the local universe
\citep{kennicutt_m101}.  The extrapolated metallicity merges into the
general IGM at radii of $\sim 80$ kpc; high redshift Damped \lya
systems generally follow this relation as well
\citep{chen_dla_abund_grad}.  BX717 also shows evidence of enhanced
metallicity, though at larger impact parameter the effect is not as
strong.


This suggests that local metal enrichment does affect the chemistry of
the IGM in the immediate surroundings of high redshift galaxies, and
that the metal-rich debris can travel distances of $100 \lesssim \rho
\lesssim 200 \hinv$ kpc before being substantially diluted by
intergalactic matter.  This encompasses an appreciable portion of the
gas structures where the galaxies reside ($\rho \lesssim 300\hinv$
kpc), but not the entire volume.  However, in cosmological
simulations, overdense filamentary structures only occupy $\sim 1-5\%$
of the total volume \citep{miralda_forest}, so the cosmic filling
factor of the outflowing material is still probably quite small unless
it escapes preferentially into voids where it could remain undetected
(see also Section \ref{sec:context}).

The ejection of interstellar heavy elements could be accomplished via
supernova-driven winds associated with star formation, or it could be
the result of tidal interactions and ram pressure stripping during the
galaxy assembly process.  Observationally it will be difficult to
distinguish between these two scenarios (or both effects working in
concert), since all are manifested by shock-heated, metal rich regions
in the vicinity of massive galaxies.  There is independent support for
the superwind hypothesis from the outflows observed in galaxy spectra
\citep{franx_winds, lbg_winds}, but the energy budget required to
drive winds from massive galaxies is quite large
\citep{maclow_ferrara}.  Alternatively, since the observed $z\sim 2.3$
galaxies have characteristic total masses of $\sim 10^{12}M_\sun$
\citep{adelberger_clustering} they must have undergone several major
mergers if the hierarchical galaxy formation paradigm is correct.
\citet{gnedin_mergers} has argued that dynamical interactions
associated with the merger process can account for most of the
intergalactic metal enrichment, but other simulations show exactly the
opposite result---that tidal disruption contributes negligibly to
intergalactic abundances, and supernova ejecta dominate
\citep{aguirre_z3_winds}.

Even if recent, local metal enrichment is important at $\rho \sim 100$
kpc, it is not clear that the heavy elements at $\rho \sim
200-300\hinv$ kpc were deposited by these same late-time processes.
Theoretical arguments based on supernova energeticsa
\citep{fujita_feedback_simulations, bruscoli_feedback, juna_feedback,
benson_feedback} and travel times \citep{aguirre_outflows,
aguirre_z3_winds} all stress the difficulty of blowing superwinds at
$z\lesssim 5$ much farther than $100$ kpc.  Particularly for large
galaxies with $M\sim 10^{11-12}M_\sun$, it appears difficult for
supernovae alone to propel substantial portions of ISM past the escape
velocity, though it may be easier to preferentially remove metal-rich
material \citep{maclow_ferrara}.  In addition, ultrapowerful winds
could create large-scale hydrodynamic disturbances in the \lya forest,
in contradiction with observations of a quiescent intergalactic
velocity field \citep{rauch_forest_hubble}.


Nevertheless, outflows are seen directly in high redshift galaxy
spectra---even for systems with $M_{\rm halo} \sim 10^{11-12}$
\citep{steidel_desert, adelberger_clustering,lbg_winds,franx_winds}.
The ejecta travel at $v\sim 250$ \kms, and their deep absorption
troughs imply large covering factors, such that the wind material is
at least $\ge 1-2$ half-light radii from the galaxy center.  If these
velocities and radii represent the true initial conditions, then most
calculations find terminal shell radii of $\sim 100$ kpc
\citep{aguirre_z3_winds, furlanetto_winds}, with the limit determined
by gravity, ram pressure, and travel time.  On the other hand, the
observable wind material could trace matter at larger radii whose
velocity has already been attenuated from a faster starting speed.  Or,
the visible low-ionization lines may only trace one phase of the total
wind, which is dominated by hot gas moving at $v\gtrsim 1000$ \kms.

In the absence of better constrained initial wind parameters, it will
remain difficult to determine beyond reasonable doubt whether the
metals at $\rho \gtrsim 200\hinv$ kpc originated in late-time winds
\citep{kurt_winds}, or whether they were already present in the IGM
from earlier times \citep[e.g.,][]{madau_ferrara_rees}.  Our
observations support the claim that absorption within
$\rho=100-200\hinv$ kpc of galaxies provides an snapshot of ongoing
galaxy feedback.  Indeed, MD103 is a clear example of a massive ($\sim
10^{12}M_\sun$) galaxy affecting the IGM as far away as $R\sim 100$
kpc.  If the outflows reach even larger radii, it may pose a challenge
for theoretical wind models, as well as measurements of \lya forest
turbulence and estimates of early metal enrichment in the IGM
\citep{pettini_z5_civ,songaila_omegaz}.



\subsection{The Chemical Dilution of Feedback Debris}\label{sec:chemical_dilution}

Consider a shell with mass $M_{\rm ej}$ and initial metallicity
$Z_{\rm ej}$ that is ejected into the IGM via supernovae or a
dynamical disruption.  Over time the shell will sweep up metal-poor
ambient matter, diluting its chemical content.  The average
metallicity of the shell may be expressed as:
\begin{eqnarray}
\left<{Z}\right> &=& {{Z_{\rm ej} M_{\rm ej} + Z_{\rm IGM} M_{\rm swept}}\over{M_{\rm ej}+M_{\rm swept}}}\\
            &\approx& Z_{\rm ej} \left(1+{{M_{\rm swept}}\over{M_{\rm ej}}}\right)^{-1}.\label{eqn:dilution}
\end{eqnarray}
The approximation in Equation \ref{eqn:dilution} refers to early times
when $Z_{\rm ej} \gg Z_{\rm IGM}$ and $M_{\rm ej} \gg M_{\rm swept}$.
It underscores how chemical dilution is only significant when the
wind's swept-up mass exceeds the mass of the initial ejecta.

For a given redshift, the expansion radius where this transition
occurs depends on the choice of $M_{\rm ej}$, which is not well
constrained on either theoretical or observational grounds.  One guess
motivated by observations of local starbursts is $M_{\rm ej}\approx
M_*$, since these systems' mass outflow rates are comparable to their
star formation rates \citep{crystal_feedback}.  However, high
abundances have been measured in the interstellar media of high
redshift galaxies \citep{alice_K20_metallicity}, so a substantial
fraction of their metals must have been retained over time.  If we
assume mass loss fractions of $[0.01, 0.1]M_*$, a fiducial galaxy
stellar mass of $M_*\sim 2\times 10^{10}M_\sun$ \citep{alice_spitzer},
and an ambient density of $\oden=10$, we find that the debris must
travel $R\sim [55,120]$ kpc before its metallicity is reduced by 1
dex.

Thus if interstellar material can escape dynamically from massive
galaxies at $z\sim 2.5-4$, it could plausibly travel more than
$R\gtrsim 100$ kpc before blending into the background metallicity
field.  Once the total shell mass becomes dominated by entrained
material, its metallicity will decrease as $R_{\rm shell}^{-3}$, or
roughly 1 dex for every factor of 2 in radius.  Accordingly the metals
would begin to blend into the surrounding IGM at $R\sim 100-200$ kpc,
much like what is shown in Figure \ref{fig:impact_metals}.  The
scatter about this simple scaling may be very large for individual
galaxies, since the particular conditions (galaxy mass, merger
history, supernova rate, mass loss, ejection velocity) vary widely
from system to system.  However, with a statistical sample over many
sightlines one could test this generic mixing prescription and in
principle measure the heavy element yield of early galaxies.


\subsection{Galaxy Formation Feedback in a Cosmological Context}\label{sec:context}

If the metal-rich absorbers indeed penetrate feedback mixing zones, it
is natural to ask what portion of the IGM is affected by the process.
Absorption line samples are often useful for addressing such problems
since they are cross-section selected and therefore relatively
unbiased with respect to luminosity.  For a galactic wind the system
cross-section is time-dependent, and represents a complicated
interplay between the starburst energetics, the progenitor's halo
mass, and the shell's entrainment and mixing with surrounding matter.

In this context, we remind the reader that these absorbers represent a
small fraction of all \lya forest lines, or even \civ systems in
general.  A cursory examination of the total \civ region in our HS1700
spectrum yielded 16 \civ detections, so the metal-rich systems
probably encompass only $\sim \frac{1}{4}-\frac{1}{3}$ of all \civ
absorbers (depending upon whether one counts the systems at $z=2.37$
and $z=2.568$ as ``metal-rich'').  The remaining 60-70\% of \civ
systems have lower \civ/\hi ratios typical of the tenuous IGM, where
$[C/H]\sim -2.8$ \citep{simcoe2004, schaye_civ_pixels, songaila_civ}.
Likewise, over the same redshift range there should be $\sim 160$
detectable \hi lines with $N_{\mhi}\ge 10^{12.5}$ \citep{kim_forest}.

One can phrase this point another way by calculating the mass fraction
of the \lya forest that is encompassed by our sample of metal-rich
feedback systems.  Since we have already calculated ionization
corrected gas densities and sizes for the absorbers, we can use the
values from Table 3 to estimate the mass density of feedback zones:
\begin{equation}
\Omega_{\rm feedback}=\left({{1}\over{\rho_c}}\right)\cdot \mu m_H \cdot
{{\sum{n_H\Delta L}}\over{\frac{c}{H_0}\sum\Delta X}}.
\end{equation}
Summing over all components from the metal-rich absorption systems
($z=1.846,2.315,$ and $2.43$) and excluding colisionally ionized \ovi
components, we find $\Omega_{\rm feedback} = 0.00061$.  We can account
statistically for the additional mass at $T\sim 10^{5-6}$ K using \ovi
measurements from \citet{simcoe2002}, who find $\Omega_\movi\sim
0.00032 (Z/0.5 Z_\sun)^{-1}(f_\movi/0.2)^{-1}$, i.e. a contribution
comparable to the cooler feedback material.  The \lya forest
encompasses $\sim 90\%$ of all baryons at $z\sim 2.3$
\citep{rauch_omegab,weinberg_omegab}, so for an assumed $\Omega_b
h^2=0.024$ \citep{wmap_params,omeara_bbn} and equal amounts of \ovi
and cooler feedback material we find:
\begin{equation}
{{\Omega_{\rm feedback}}\over{\Omega_{\rm forest}}} \sim 3\%.
\end{equation}
Evidently the shock-heated, high-metallicity regions (i.e. the systems
most clearly related to feedback) only represent a small fraction of
the total mass in the \lya forest, unless their mass is dominated by a
very hot phase with $T\gtrsim 10^{6-7}$ K.  Gas at these temperatures
is seen in X-ray observations of low redshift starbursts, though it
would be nearly impossible to detect at $z\sim 2.3$.  Also, if the
cooler, lower metallicity \civ systems at larger radii are tepid
remnants of powerful winds at $z\sim 2-4$ the mass fraction of
feedback systems could increase by a factor of a few.  Even then the
majority of baryons are still found at densities near the cosmic mean,
where individual \civ systems cannot be measured for cross-correlation
with galaxies.

While the metal-rich systems comprise a small fraction of all \lya and
\civ systems by {\em number}, they may represent a significant amount
of the total \civ {\em mass} in the universe.  This can be seen by
examining the customary formula for calculating the contribution of
\civ to closure density \citep{lanzetta_omega_civ}:
\begin{equation}
\Omega_{\mciv}={{H_0 m_{\mciv}}\over{c\rho_{\rm crit}}}\int_{N_{\rm
min}}^{N_{\rm max}}Nf(N)dN
\end{equation}
where $f(N)$ represents the number of \civ systems per unit column
density per unit absorption pathlength, and $N_{\rm min}$ and $N_{\rm
max}$ represent the minimum and maximum \civ mass in the absorption
line sample.  \citet{songaila_omegaz} finds a power-law form for
$f(N)$ with slope $f(N)\propto N^{-1.8}$ at $z=2.90-3.54$, which
implies that $\Omega_\mciv\propto (N_{\rm max}^{0.2}-N_{\rm
min}^{0.2})$.  Accordingly, \civ mass density calculations are always
dominated by the few highest column density systems in a sample, which
are heavily represented among the metal-rich systems studied here.  In
fact, the systems chosen for our modeling analysis are the six
strongest \civ absorbers of the 16 total toward HS1700 (recall that
they were not selected according to $N_\mciv$); all have at least one
component with $N_\mciv>10^{13}$.  Of the remaining 10 \civ systems,
only one has $N_\mciv=10^{13.007}$; all the rest have $N_\mciv \le
10^{13}$.

This is an important consideration for studies of early metal
enrichment in the IGM via measurements of $\Omega_\mciv$
\citep{pettini_z5_civ,songaila_omegaz}.  Since high-redshift
($z\gtrsim 5$) QSO observations are often undertaken at lower spectral
resolution and/or signal-to-noise ratios, the line samples used for
these measurements may be disproportionately populated with strong
\civ systems.  If these trace dense, metal-rich structures like the
ones presented here, the $\Omega_\mciv$ measurements may not in fact
be tracing chemical enrichment in tenuous regions of the IGM.
Instead, the ionization and chemistry may be governed more locally,
e.g. by galaxies and feedback \citep[as suggested in][Section
6]{pettini_z5_civ}.  This provides a possible explanation of the
observed lack of evolution in $\Omega_\mciv$ with redshift: at each
epoch the strongest \civ lines may provide an instantaneous snapshot
of regions which are actively undergoing chemical enrichment and being
illuminated by local stars, rather than an integral of global
enrichment in the background-illuminated ``field'' \civ systems over
cosmic time.

\section{Summary and Conclusions}\label{sec:conclusions}

We have presented observations of galaxies and intergalactic gas
toward the $z=2.73$ quasar HS1700+6416, with the goal exploring the
effects of galaxy formation feedback on the IGM.  The unique aspect of
our analysis is its careful treatment of the absorption systems---we
have performed detailed line fits and full ionization simulations to
determine metallicites, gas densities, sizes, and [Si/C] relative
abundances in the IGM.  Our galaxy and absorber samples are still
quite small and subject to the associated small-number caveats: we
have considered six absorption systems identified by the presence of
\ovi, \nv, or \mgii in a single quasar sightline, together with 14
galaxies located within $500\hinv$ kpc of the sightline.  However, the
data quality is uniformly excellent, and even in large surveys the
total number of galaxies is not as important as the density of objects
close to the quasar sightline ($\lesssim 40\arcsec$).  The main
results of our analysis may be summarized as follows:
\begin{enumerate}
\item{Our absorption selection identifies regions of high metallicity
in the IGM, with half of the six systems exhibiting near-solar values.
The metal-rich systems have small absorption thickness ($\Delta
L\lesssim 1$ kpc) and high gas density ($\oden \gtrsim 100$), and are
typically mixed with shock-heated \ovi.  The remaining systems
resemble moderately overdense intergalactic filaments with slightly
higher than normal chemical enrichment.}
\item{In regions where we could measure [Si/C] relative abundances,
there are indications of a 0.1-0.5 dex silicon enhancement.  This may
indicate a preferential enrichment by debris from Type II supernovae.}
\item{Luminous galaxies are found near two of the metal-rich absorbers
at identical redshift and impact parameter $\rho \lesssim 200\hinv$
physical kpc.  The strongest absorber in the sample is associated with
the closest galaxy to the QSO sightline.}
\item{The absorption systems most likely arise in thin sheet- or
shell-like structures.  This provides an efficient way to produce
large absorption cross-sections for structures with small linear
dimensions.  If the shells form as bubbles around high redshift
galaxies, then the bubbles have radii of $R\gtrsim 100$ kpc and
thickness $\sim 1$ kpc.  Roughly half of these bubbles may be
associated with known luminous galaxies, while the other half may come
from galaxies with either slightly lower luminosities, or different
colors than current $z\sim 2.3$ samples.}
\item{A generic model of radiatively efficient shocks plowing into
modestly overdense filaments ($\oden\sim 10$) explains the basic
observed properties of the absorption systems.}
\item{We see evidence for a distinct dropoff in the intergalactic gas
density (traced by $\nhi$) at impact parameters of $\sim 1$ comoving
Mpc from galaxies, in agreement with the results of
\citet{adelberger_z2}.  At this transition, the baryonic density
declines from $\oden \sim 10-100$ to values at or slightly below the
mean.  The galaxies appear to be embedded in intergalactic gas
structures with this scale on the sky.}
\item{The metallicity field is strongly enhanced within $\rho\lesssim
100-200\hinv$ physical kpc of galaxies, which we interpret as a
signature of galaxy-formation feedback.  This feedback could be
triggered by supernova-driven winds, or by the violent disruption of a
proto-galaxy's ISM during hierarchical mergers.  At larger radii the
metallicity field resembles that of the general IGM, with abundances
of $[X/H]\lesssim -2$.  We showed that metal-rich debris originating
in galaxies could easily travel $\sim 100$ kpc before sweeping up
enough metal-poor ambient matter to dilute into the background
metallicity field.}
\item{Active (i.e. shock-heated and undiluted) feedback zones only
comprise a few percent of the total mass of baryons in the \lya
forest.  However, they are heavily represented in the strongest \civ
systems and may comprise a substantial fraction of the \civ mass in
the universe.  This is significant for enrichment studies at high
redshift ($z\gtrsim 5$) that make use of $\Omega_\mciv$.}
\end{enumerate}

It appears that star and/or galaxy formation feedback does
significantly affect the properties of the IGM within $R\sim
100-200\hinv$ kpc of $z\sim 2.3$ galaxies.  It is less clear whether
the much more dilute heavy elements observed in the widespread IGM
originated from these same processes.  Powerful outflows are required
to cover distances $\gtrsim 150$ kpc, which is somewhat uncomfortable
for theoretical feedback models and observations of the \lya forest
velocity field.  Yet the correlation function between galaxies and
\civ systems extends to larger scales than what we observe for our
metal-rich feedback systems.  This fact has been used to argue for a
combination of more energetic winds at small scales and clustering at
large scales \citep{kurt_winds, adelberger_z2}, or early chemical
pollution of the biased regions where massive galaxies form
\citep{porciani_madau_lbg_metals, scannapieco_metal_enrichment}.  It
remains to be seen whether these discrepancies can be resolved by
better wind models, or whether the large-scale correlations simply
reflect that galaxies are embedded in intergalacic structures that
were pre-enriched at $z\sim 6-10$.

\acknowledgements

We are grateful to C. Steidel and team for providing us with
information about galaxy redshifts near HS1700+6416 in advance of
publication, and for helpful comments on a draft of the manuscript.
We also thank H.W. Chen for helpful discussions.  R.S. acknowledges
financial support from the MIT Pappalardo Fellowship program, and from
an AAS Small Research grant.  W.L.W.S thanks the National Science
Foundation for supporting this work under grant AST 02-06067, and
M.R. also thanks the NSF for support under grant AST 00-98492.
Finally, we extend our thanks to those of Hawaiian ancestry who
generously share their sacred mountain for the advancement of
astronomical research.

\clearpage

\bibliography{../ovi}

\begin{deluxetable}{c c c c c c c c}
\tablewidth{0pc} \tablecaption{Confirmed and Candidate $z\sim 2-2.5$
Galaxies near HS1700+6416} 

\tablehead{{Galaxy} & {$\rho$($h_{71}^{-1}$ kpc)} & {$z$} & {$R$} &
{$M_{\rm 2000\AA}$} & {$L/L^{*}$} & {$\log(M_*/M_\sun)$} & {SFR
($M_\sun$ yr$^{-1}$)}}
 
\startdata 
MD103\dagnote  & 115.3 & 2.3148 &  24.23 & -22.09 & 0.81 & 11.07 & 64 \\
MD109\dagnote  & 208.0 & 2.2942 &  25.46 & -20.84 & 0.26 & 10.48 & 14 \\
BX717\dagnote  & 216.3 & 2.4353 &  24.78 & -21.66 & 0.49 & 9.85  & 20 \\
BX691\dagnote  & 290.1 & 2.1895 &  25.33 & -20.84 & 0.29 & 11.04 & 33 \\
BX709   & 316.1 & 2.285  &  25.18 & -21.11 & 0.34 & 10.37 & 10 \\
BX635   & 327.3 & 1.860  &  24.87 & -20.86 & 0.45 & \nodata & \nodata \\
MD119   & 336.2 & 2.566  &  25.04 & -21.44 & 0.38 & \nodata & \nodata \\
BX759   & 369.9 & 2.418  &  24.43 & -21.99 & 0.67 & 10.52 & 73 \\
MD92    & 431.1 & 2.691  &  25.47 & -21.04 & 0.26 & \nodata & \nodata \\
BX756   & 473.7 & 1.738  &  23.21 & -22.34 & 2.07 & 9.76 & 377 \\
\hline
BX767   & 409.0 &        &  24.64 & \\
BX629   & 410.0 &        &  24.06 & \\
BX720   & 421.9 &        &  24.74 & \\
BX632   & 428.2 &        &  25.12 & \\

\enddata 
\tablenotetext{$\dagger$}{Redshift and SFR determined from
H$\alpha$ emission line \citep{erb_halpha}.  Other objects' redshifts
were determined from optical spectra \citep{steidel_desert}, and SFR
from population synthesis models \citep{alice_spitzer}.}
 
\end{deluxetable}
\label{tbl:galaxies}
\clearpage
\LongTables
\begin{longtable}{l l l l l l | l l}
\tablewidth{0pc}
\tablecaption{Voigt Profile Fit Components}
\tablehead{ & \multicolumn{3}{c}{\em Voigt Profile Parameters} & \multicolumn{2}{c}{\em Linewidth $T$} & \multicolumn{2}{c}{\em Model Params} \\
{Ion} & {$z$} & {$b$ (km s$^{-1}$)} & {$\log N$ (cm$^{-2}$)} & {$T_{b}$ (K)} & {$b_{nt}$} & {$\log N_{\rm mod}$} & {$T_{eq}$}}
 
\startdata
\hline
\multicolumn{8}{c}{Voigt Profile Components: $z=1.845$ absorption system}\\
\hline
  \siiv  &  $1.844840\pm0.000002$ & $3.14\pm0.19$     & $12.498\pm0.030$ & $\le 16,000$ & \nodata & 12.465 & 10,000 \\
  \siiii &                        & $3.14\dagnote$    & $12.452\pm0.094$ &           &         & 12.505 \\
  \siii  &                        & $3.14\dagnote$    & $\le 11.706$     &           &         & 11.221 \\
  \cii   &                        & $4.81\dagnote$    & $12.631\pm0.098$ &           &         & 12.615 \\ 
  \civ   &                        & $4.81\dagnote$    & $13.490\pm0.023$ &           &         & 13.492 \\
  \hi    &                        & $23.00\dagnote$   & $<16.0$          &           &         & 14.641 \\

\hline
  \siiv  &  $1.845035\pm0.000001$ & $ 3.52\pm0.43 $   & $13.001\pm0.061$ & $\le 26,000$ & \nodata & 12.979 & 8,200 \\
  \siii  &                        & $ 3.52\dagnote$   & $12.592\pm0.025$ &           &         & 12.560 \\
  \siiii &                        & $ 3.52\dagnote$   & $13.512\pm0.156$ &           &         & 13.369 \\
  \cii   &                        & $ 6.09\pm0.39 $   & $13.692\pm0.020$ &           &         & 13.613 \\
  \alii  &                        & $ 6.08\pm1.87 $   & $11.041\pm0.078$ &           &         & 11.610 \\
  \mgii  &                        & $ 3.09\pm0.18 $   & $12.442\pm0.016$ &           &         & 12.437 \\
  \hi    &                        & $20.00\dagnote$   & $< 16.5 $        &           &         & 15.580 \\

\hline	
  \nv    &  $1.845044\pm0.000002$ &   $15.42\pm3.50$ &  $13.421\pm0.091$ & $\lesssim200,000$ & $\le 13.7$  & 13.324 & $\gtrsim 14,000$ \\
  \siiv  &                        &   $14.59\pm1.16$ &  $12.951\pm0.025$ &           &         & 13.026  \\
  \civ   &                        &   $16.11\dagnote$&  $14.430\pm0.007$ &           &         & 14.472  \\
  \hi    &                        &   $35.0\tablenotemark{a}$ &  $\lesssim15.5$&     &         & 14.619 \\

\hline		    	 		 	 	 
  \siiv  &  $1.845410\pm0.000010$ &   $5.88\pm1.34$  &  $12.413\pm0.100$ & $\lesssim 60,000$ & $\sim 0$ & 12.427 & 10,000 \\
  \siiii &                        &   $5.88\dagnote$ &  $12.369\pm0.036$ &           &         & 12.376 \\
  \civ   &                        &   $9.11\pm0.83$  &  $13.669\pm0.064$ &           &         & 13.648 \\

\hline	
  \siiv  &  $1.845511\pm0.000016$ &   $4.34\pm2.97$  &  $12.120\pm0.183$ & $\lesssim 35,000$ & \nodata & 12.128 & 8,400\\
  \siiii &                        &   $4.34\dagnote$ &  $12.490\pm0.050$ &           &         & 12.468 \\
  \civ   &                        &   $7.35\pm1.62$  &  $13.012\pm0.267$ &           &         & 13.031 \\

\hline
  \nv   &  $1.845258\pm0.000130$ &  $24.44\pm11.95$ &  $13.091\pm0.305$  & & & \\
\hline
\hline
\multicolumn{8}{c}{Voigt Profile Components: $z=2.168$ absorption system}\\
\hline
\hline
  \civ   &  $2.167537\pm0.000008$ &  $12.89\pm1.15$  & $12.808\pm0.030$ & $<120,000$ & \nodata &\nodata & \nodata \\
  \hi    &                        &  $30.0\dagnote$  & $\lesssim 14.5$  & & &$\le 14.5$ \\
\hline  
  \siiv  &  $2.167935\pm0.000002$ &  $11.15\dagnote$ & $13.159\pm0.010$ & $\le 70,000$ & $\ge 9.0$ & 13.104 & 29,900\\
  \siiii &                        &  $11.15\pm0.21$  & $13.021\pm0.019$ & & & 13.304 \\
  \cii   &                        &  $13.44\dagnote$ & $12.989\pm0.032$ & & & 12.734 \\
  \civ   &                        &  $13.44\pm0.20$  & $13.987\pm0.010$ & & & 13.995 \\
  \mgii  &                        &  $ 8.21\pm4.28$  & $11.391\pm0.153$ & & & 11.625 \\
  \alii  &                        &  $10.04\pm9.22$  & $10.801\pm0.397$ & & & 10.808 \\
  \hi    &			  &  $25.0\tablenotemark{b}$ & $\lesssim16.8\tablenotemark{b}$        & & & 16.368 \\
\hline
  \siiv  &  $2.168064\pm0.000003$ &  $ 4.90\pm0.41$  & $12.469\pm0.044$ & $\le 15,000$ & \nodata & 12.518 & 22,000 \\
  \siiii &                        &  $ 4.90\dagnote$ & $13.218\pm0.106$ & & & 13.067 \\
  \cii   &		          &  $ 4.35\pm1.59$  & $12.331\pm0.109$ & & & 12.326 \\
  \civ   &			  &  $ 4.35\dagnote$ & $12.564\pm0.171$ & & & 12.763 \\
  \mgii  &			  &  $ 3.26\pm2.06$  & $11.401\pm0.113$ & & & 11.599 \\
  \alii  &			  &  $23.32\pm18.26$ & $10.956\pm0.391$ & & & 10.804 \\
  \hi    &			  &  $25.0\tablenotemark{b}$ & $\lesssim16.2\tablenotemark{b}$        & & & 16.246 \\
\hline
  \hi    &  $2.168938\pm0.000009$ &  $25.58\pm0.80$ & $13.870\pm0.016$ \\
\hline
\hline
\multicolumn{8}{c}{Voigt Profile Components: $z=2.315$ absorption system}\\
\hline
\hline
  \civ  &   $2.313843\pm0.000012$  & $8.14\pm0.15$  & $12.392\pm0.081$ & $<47,000$ & \nodata & 12.392 & 6600\\
  \hi   &                          & $28.10\dagnote$& $14.225\pm0.028$ & & & 14.214 \\
\hline
  \siii &   $2.314895\pm0.000016$ &  $ 7.52\pm1.03$ & $12.660 \pm 0.086$ & $\le 90,000$ & $\ge 0$ & 12.580 & $5400$ \\
  \nii  &                         &  10.64\dagnote  & $12.592 \pm 0.120$ & & & 12.585 \\
  \alii &                         &   7.67\dagnote  & $11.116 \pm 0.108$ & & & 11.390 \\
  \mgii &                         &   8.00\dagnote  & $12.324 \pm 0.091$ & & & 12.292 \\
  \cii  &                         &  11.50\dagnote  & $13.420 \pm 0.105$ & & & 13.350 \\
\hline		  			  	  	  
  \siii &   $2.314996\pm0.000002$ &  $ 3.31\pm0.32$ & $12.831 \pm 0.058$ & $\le 18,000$ & $>0$ & 12.825 & $\le5400$ \\
  \nii  &                         &   4.68\dagnote  & $12.724 \pm 0.069$ & & & 12.882 \\
  \alii &                         &   3.38\dagnote  & $11.261 \pm 0.073$ & & & 11.463 \\
  \feii &                         &   2.35\dagnote  & $11.606 \pm 0.106$ & & & 12.001 \\
  \mgii &                         &   3.56\dagnote  & $12.705 \pm 0.036$ & & & 12.716  \\
  \cii  &                         &   5.06\dagnote  & $13.648 \pm 0.061$ & & & 13.436 \\
  \hi   &                         &  18.51\dagnote  & $15.5\lesssim \nhi \le 16.862$  & & & 15.576 \\
\hline		  			  	  	  
  \siii &   $2.315504\pm0.000001$ &  $ 3.01\pm0.09$ & $12.943 \pm 0.022$ & $\le 15,000$ & $>0$ & 12.858 & $\le 5400$ \\
  \nii  &			  &   4.25\dagnote  & $13.107 \pm 0.027$ & & & 13.150 \\
  \cii  &			  &   4.60\dagnote  & $13.643 \pm 0.008$ & & & 13.687 \\
  \mgii &			  &   3.23\dagnote  & $12.579 \pm 0.019$ & & & 12.653 \\
  \feii &			  &   2.13\dagnote  & $12.119 \pm 0.025$ & & & 12.338 \\
  \alii &			  &   3.07\dagnote  & $11.463 \pm 0.045$ & & & 11.727 \\
  \hi   &                         &  15.90\dagnote  & $14.939\le \nhi \lesssim 17.0$ & & & 15.569 \\
\hline
  \siiv &   $2.316350\pm0.000001$ &  $4.08\pm0.06$  & $12.963\pm0.005 $ & $16,600$ & $2.6$ & 12.977 & 18,000\\
  \civ  &                         &  $5.46\pm1.77$  & $13.944\pm0.160 $ & & & 14.295 \\
  \nv   &                         &  $8.12\pm0.69$  & $13.198\pm0.024 $ & & & 12.997 \\
  \hi   &                         &  $21.75\dagnote$& $\lesssim 14.3  $ & & & 14.127 \\
\hline
  \siiv &   $2.314470\pm0.000007$ &  $5.19\pm0.78$  & $11.853\pm0.065$ & $\le 45,000$ & \nodata & 11.842 & 11,000 \\
  \civ  &                         &   7.93\dagnote  & $13.057\pm0.048$ & & & 13.068 \\
  \hi   &                         &   27.5\dagnote  & $\lesssim 14.7$  & & & 13.126 \\
\hline
  \ovi  &   $2.314992\pm0.000019$ &  $11.90\dagnote$& $13.951\pm0.108$ & \nodata &\nodata & \nodata & $\gtrsim 230,000$ \\
  \nv   &                         &  $12.72\pm1.41$ & $\le 13.043$     & & &  \\
\hline
  \ovi  &   $2.315182\pm0.000012$ &  $11.64\dagnote$& $14.259\pm0.054$ & \nodata &\nodata & \nodata & $\gtrsim 250,000$ \\
  \nv   &                         &  $12.44\pm1.18$ & $\le 13.026$     & & &  \\
\hline 
  \siiv&    $2.314900\pm0.000005$ &  $7.45\pm0.26$   & $13.550\pm0.036$ & $\lesssim 43,000$ & $\gtrsim 5.5$ & 13.526 & 8300\\
  \civ &                          &  $9.46\pm0.48$   & $14.309\pm0.042$ & & & 14.295 \\
  \nv  &                          &  $10.75\pm4.84$  & $12.741\pm0.152$ & & & 12.782 \\
  \hi  &                          &  $32.77\dagnote$ & $\lesssim 15.4 $ & & & 15.094 \\
\hline
  \nv  &    $2.315117\pm0.000007$ &  $11.97\pm1.68$  & $13.154\pm0.054$ & $<70,000$ & $>0$ & 13.163 & 12,000 \\
  \civ &                          &  $10.00\pm0.96$  & $14.391\pm0.047$ & & & 14.381 \\
  \siiv&                          &                  & $\lesssim 13.1$  & & & 13.010 \\
  \hi  &                          &  $34.6\dagnote$  & $\lesssim 16.1$  & & & 14.286 \\
\hline
  \nv   &   $2.315556\pm0.000010$ &  $22.57\pm2.38$  & $13.186\pm0.050$  & \nodata & \nodata & 13.137 & 14,000 \\
  \civ  &                         &  $24.38\dagnote$ & $14.420\pm0.030$  & & & 14.370 \\
  \siiv &                         &  $24.38\dagnote$ & $\lesssim 12.856$ & & & 12.976 \\
  \hi  &                          &  $30.0\dagnote$  & $\lesssim 15.7$   & & & 14.536 \\
\hline
  \civ  &   $2.315868\pm0.000003$ &   $7.52\pm0.77$   & $13.923\pm0.062$ & $\le 40,000$ & \nodata & 13.932 & 15,500 \\
  \nv   &                         &   $6.96\dagnote$  & $12.801\pm0.074$ & & & 12.787 \\
  \siiv &                         &   $4.91\dagnote$  & $12.387\pm0.031$ & & & 12.392 \\
  \hi   &                         &   $26.0\dagnote$  & $\lesssim 15.0$  & & & 13.972 \\
\hline
  \nv   &   $2.316071\pm0.000004$ &   $5.28\pm0.93$   & $12.539\pm0.088$ & $\le 23,000$ & \nodata & 12.378 & 12,500 \\
  \civ  &                         &   $5.70\dagnote$  & $13.248\pm0.057$ & & & 13.514 \\
  \siiv &                         &   $3.73\dagnote$  & $12.002\pm0.043$ & & & 11.998 \\
  \hi   &                         &   $20.0\dagnote$  & $\lesssim 14.7$  & & & 13.289 \\
\hline
  \ovi  &   $2.315790\pm0.000014$ &  $34.43\pm2.14$ & $14.675\pm0.021$ & & & \nodata & $\gtrsim 250,000$\\ 
  \nv   &                         &  $31.87\dagnote$& $\le 13.415$     & & &  & \\ 
\hline
  \ovi  &   $2.316345\pm0.000011$ &  $13.20\pm1.53$ & $14.068\pm0.065$ & & & \nodata & $\gtrsim 230,000$\\ 
  \nv   &                         &  $12.22\dagnote$& $\le 13.261$     & & &  & \\ 
\hline
  \ovi  &   $2.316693\pm0.000063$ &  $45.16\pm4.91$ & $14.087\pm0.065$ & & & \nodata & $\gtrsim 260,000$\\ 
  \nv   &                         &  $41.81\dagnote$& $\le 12.632$     & & & & \\ 
\hline
  \ovi  &   $2.314618\pm0.000068$ &  $46.80\pm14.92$& $13.829\pm0.125$ & & & \nodata & $\gtrsim 260,000$\\ 
  \nv   &                         &  $43.30\dagnote$& $\le12.612$      & & & & \\ 
\hline
  \civ  &   $2.315455\pm0.000011$ &   $1.06\pm4.64$ & $13.517\pm5.127$ & & & \nodata\\
  \civ  &   $2.316317\pm0.000013$ &  $ 8.97\pm0.44$ & $14.074\pm0.170$ & & & \nodata\\ 
  \civ  &   $2.316587\pm0.000011$ &  $ 6.14\pm1.49$ & $12.625\pm0.075$ & & & \nodata\\ 
  \siiv &   $2.315501\pm0.000003$ &   $2.00\pm1.39$ & $12.938\pm0.088$ & & & \nodata\\
  \siiv &   $2.314995\pm0.000003$ &  $ 2.76\pm0.79$ & $13.427\pm0.105$ & & & \nodata\\ 
  \siiv &   $2.315128\pm0.000003$ &  $ 6.11\pm0.79$ & $13.069\pm0.053$ & & & \nodata\\ 
  \siiv &   $2.315395\pm0.000076$ &  $13.75\pm6.60$ & $13.106\pm0.326$ & & & \nodata\\ 
  \siiv &   $2.315578\pm0.000078$ &  $ 8.93\pm7.47$ & $12.856\pm0.695$ & & & \nodata\\ 
  \siiv &   $2.315757\pm0.000007$ &  $ 2.25\pm1.99$ & $12.172\pm0.152$ & & & \nodata\\ 
\hline
\hline
\multicolumn{8}{c}{Voigt Profile Components: $z=2.379$ absorption system}\\
\hline
\hline
  \civ   & $2.379919\pm0.000002$ &   $9.92\pm0.26$  & $13.056\pm0.008$ & $50,000$ & $5.24$ &  13.043 & 29,000 \\
  \ciii  &                       &   $9.92\dagnote$ & $13.363\pm0.088$ &  & &  13.430 \\
  \siiii &                       &   $7.61\pm0.71$  & $11.755\pm0.032$ &  & &  11.874 \\
  \siiv  &                       &   $7.61\dagnote$ & $11.788\pm0.039$ &  & &  11.657 \\
  \siii  &                       &   $7.61\dagnote$ & $\le 11.610$     &  & &  10.697 \\
  \hi    &                       &   $23.84\pm0.69$ & $15.393\pm0.057$ &  & &  15.490 \\
\hline
  \ovi   & $2.379988\pm0.000009$ &   $19.57\pm1.12$ & $13.518\pm0.021$ & $\le 370,000$ & \nodata & \nodata & $\gtrsim 280,000$ \\
  \nv    &                       &   $18.1\dagnote$ & $\le 12.005$     &          & & &\\
\hline
  \hi    & $2.379291\pm0.000057$ &   $50.00\pm2.61$ & $14.487\pm0.065$ & & & &\\  
  \hi    & $2.380634\pm0.000047$ &   $26.20\pm3.52$ & $13.098\pm0.088$ & & & &\\
\hline
\hline
\multicolumn{8}{c}{Voigt Profile Components: $z=2.43$ absorption system}\\
\hline
\hline
  \cii   &  $2.433052\pm0.000001$ & $  7.37\dagnote$ & $12.756\pm    0.049$ & $17,500$ & 5.48 & 12.655 & 19,700 \\
  \ciii  &                        & $  7.37\dagnote$ & $13.884\pm    0.516$ & & & 13.887\\
  \civ   &                        & $  7.37\pm0.66$ & $12.940\pm    0.133$ & & & 12.967\\
  \siii  &                        & $  6.36\dagnote$ & $12.129\pm    0.042$ & & & 12.018\\
  \siiii &                        & $  6.36\dagnote$ & $13.134\pm    0.018$ & & & 13.156\\
  \siiv  &                        & $  6.36\dagnote$ & $12.582\pm    0.018$ & & & 12.580\\
  \alii  &                        & $  6.44\dagnote$ & $10.941\pm    0.072$ & & & 11.187\\
  \hi    &                        & $ 17.92\dagnote$ & $15.751\pm    0.250$ & & & 15.887 \\
\hline	     	  	            	  	        	  		 	  
  \cii   &  $2.433426\pm0.000008$ & $ 16.74\dagnote$ & $12.726\pm    0.053$ & $\le 60,000$ & 13.95 & 12.627 & 19,700\\
  \civ   &                        & $ 16.74\pm2.46$ & $12.650\pm    0.234$ & & & 12.939\\
  \siii  &                        & $ 15.21\dagnote$ & $11.975\pm    0.025$ & & & 11.864\\
  \siiii &                        & $ 15.21\pm0.84$ & $12.891\pm    0.027$ & & & 13.003\\
  \siiv  &                        & $ 15.21\dagnote$ & $12.495\pm    0.040$ & & & 12.427\\
  \hi    &                        & $ 34.95\dagnote$ & $15.719\pm    0.354$ & & & 15.859\\
\hline	     	  	            	  	        	  		 	  
  \cii   &  $2.433650\pm0.000001$ & $ 10.54\dagnote$ & $12.446\pm    0.071$ & $\le 55,000$ & $\ge 5.93$ & 12.338 & 19,700\\
  \civ   &                        & $ 10.54\pm0.84$ & $12.643\pm    0.050$ & & & 12.650\\
  \siii  &                        & $  8.23\dagnote$ & $11.371\pm    0.071$ & & & 11.626\\
  \siiii &                        & $  8.23\pm0.72$ & $12.751\pm    0.048$ & & & 12.764\\
  \alii  &                        & $  8.54\dagnote$ & $10.758\pm    0.094$ & & & 10.870\\
  \hi    &                        & $ 30.76\dagnote$ & $15.641\pm    0.153$ & & & 15.570\\
\hline	     	  	            	  	        	  		 	  
  \civ   &  $2.431814\pm0.000012$ & $  7.22\pm1.78$ & $12.273\pm    0.123$ & \nodata & \nodata & 12.162 & 35,000\\
  \ciii  &                        & $  7.22\dagnote$ & $12.145\pm    0.228$ & & & 12.294\\
  \hi    &                        & $ 25.01\dagnote$ & $15.055\pm    0.022$ & & & 14.861\\
\hline	     	  	            	  	        	  		 	  
  \civ   &  $2.432091\pm0.000015$ & $ 18.20\dagnote$ & $12.841\pm    0.041$ & \nodata & \nodata & 12.818 & 31,800\\
  \ciii  &                        & $ 18.20\pm2.05$ & $12.998\pm    0.047$ & & & 13.023\\
  \hi    &                        & $ 29.84\pm1.20$ & $15.401\pm    0.069$ & & & 15.370\\
\hline	     	  	            	  	        	  		 	  
  \civ   &  $2.432632\pm0.000017$ & $ 11.82\pm1.84$ & $12.535\pm    0.074$ & \nodata & \nodata & \\
  \ciii  &                        & $ 11.82\dagnote$ & $12.970\pm    0.070$ & & & \\
  \siiv  &                        & $ 50.00\pm37.76$ & $12.076\pm    0.278$ & & & \\
\hline	     	  	            	  	        	  		 	  
  \civ   &  $2.432933\pm0.000026$ & $ 10.97\pm2.71$ & $12.927\pm    0.139$ & \nodata & \nodata & 12.882 & 24,000\\
  \ciii  &                        & $ 10.97\dagnote$ & $13.227\pm    0.181$ & & & 13.504\\
  \cii   &                        & $ 10.97\dagnote$ & $12.183\pm    0.168$ & & & 12.048\\
  \hi    &                        & $ 26.01\pm5.88$ & $15.567\pm    0.147$ & & & 15.423 \\
\hline	     	  	            	  	        	  		 	  
  \ovi   &  $2.432600\pm0.000145$ & $ 42.97\pm12.70$ & $13.467\pm    0.164$ & $\le 1,700,000$ & \nodata & \nodata & $\gtrsim 224,000$ \\
  \nv    &                        & $ 42.97\dagnote$ & $12.788\pm    0.082$ & & & \\
\hline	     	  	            	  	        	  		 	  
  \ovi   &  $2.433233\pm0.000030$ & $ 26.64\pm5.63$ & $13.590\pm    0.141$ & $\le 680,000$ & \nodata & \nodata & $\gtrsim 280,000$\\
  \nv    &                        & $ 26.60\dagnote$ & $\le 12.030 $     & & & \\
\hline	     	  	            	  	        	  		 	  
  \ovi   &  $2.433749\pm0.000037$ & $ 18.15\pm3.72$ & $13.158\pm    0.116$ & $\le 316,000$ & \nodata & \nodata & $\gtrsim 280,000$ \\
  \nv    &                        & $ 18.15\dagnote$ & $11.531\pm    0.213$ & & & \\
\hline	     	  	            	  	        	  		 	  
  \civ   &  $2.433285\pm0.000021$ & $ 13.86\pm3.59$ & $12.876\pm    0.160$ & & & \\
\hline	     	  	            	  	        	  		 	  
  \cii   &  $2.438637\pm0.000003$ & $ 10.92\pm0.49$ & $12.001\pm    0.094$ & $\le 85,000$ & \nodata & 11.872 & 28,500 \\
  \ciii  &                        & $ 10.92\dagnote$ & $13.492\pm    0.052$ & & & 13.458\\
  \civ   &                        & $ 10.92\dagnote$ & $13.079\pm    0.016$ & & & 13.080\\
  \siii  &                        & $  7.14\dagnote$ & $\le11.500$       & & & \\
  \siiii &                        & $  7.14\dagnote$ & $12.111\pm    0.032$ & & & 12.218\\
  \siiv  &                        & $  7.14\dagnote$ & $12.259\pm    0.025$ & & & 12.430\\
  \hi    &                        & $ 37.83\dagnote$ & $15.510\pm    0.035$ & & & 15.405 \\
\hline	     	  	            	  	        	  		 	  
  \cii   &  $2.438857\pm0.000002$ & $  6.55\pm0.46$ & $11.946\pm    0.087$ & 21,000 & 3.75 & 12.064 & 22,000 \\
  \ciii  &                        & $  6.55\dagnote$ & $13.626\pm    0.126$ & & & 13.517\\
  \civ   &                        & $  6.55\dagnote$ & $12.848\pm    0.022$ & & & 12.864\\
  \siii  &                        & $  5.14\dagnote$ & $\le 11.577 $        & & & \\
  \siiii &                        & $  5.14\pm0.41$ & $12.248\pm    0.040$ & & & 12.452\\
  \siiv  &                        & $  5.14\dagnote$ & $12.200\pm    0.025$ & & & 12.065\\
  \hi    &                        & $ 19.00\dagnote$ & $15.316\pm    0.084$ & & & 15.240\\
\hline	     	  	            	  	        	  		 	  
  \siiii &  $2.438286\pm0.000006$ & $ 21.61\pm2.28$ & $12.147\pm    0.034$ & $\le 100,000$ & \nodata & 12.173 & 31,800 \\
  \siiv  &                        & $ 21.61\dagnote$ & $12.021\pm    0.062$ & & & 12.022\\
  \ciii  &                        & $ 11.17\dagnote$ & $12.963\pm    0.041$ & & & 12.952\\
  \civ   &                        & $ 11.17\pm0.86$ & $12.671\pm    0.028$ & & & 12.656\\
  \hi    &                        & $ 13.15\pm0.96$ & $15.221\pm    0.089$ & & & 15.086\\
\hline	     	  	            	  	        	  		 	  
  \ciii  &  $2.437774\pm0.000004$ & $  8.53\pm0.55$ & $12.835\pm    0.043$ & $\le 52,000$ & \nodata & 12.796 & 27,000\\
  \civ   &                        & $  8.53\dagnote$ & $12.682\pm    0.022$ & & & 12.636\\
  \siiii &                        & $  5.58\dagnote$ & $11.196\pm    0.111$ & & & 11.321\\
  \siiv  &                        & $  5.58\dagnote$ & $11.331\pm    0.112$ & & & 11.310\\
  \hi    &                        & $ 31.75\pm0.52$ & $13.992\pm    0.014$ & & & 13.960\\
\hline	     	  	            	  	        	  		 	  
  \civ   &  $2.440517\pm0.000005$ & $  9.80\pm0.54$ & $13.080\pm    0.028$ & $\le 70,000$ & \nodata & 12.782 & 27,000 \\
  \siiv  &                        & $  9.80\dagnote$& $\le 11.596$         & & & 11.422\\
  \hi    &                        & $ 34.08\dagnote$& $14.558\pm    0.022$ & & & 14.536\\
\hline	     	  	            	  	        	  		 	  
  \ovi   &  $2.439748\pm0.000013$ & $  14.8\pm1.86$ &  $13.570\pm    0.044$ & $\lesssim 210,000$ & \nodata & \nodata &$\gtrsim 234,000$\\
  \nv    &                        & $  14.8\dagnote$ &  $\le 12.533  $    & & & \\
\hline	     	  	            	  	        	  		 	  
  \ovi   &  $2.440253\pm0.000032$ & $ 15.96\pm3.90$  &  $13.492\pm   0.137$ & $\lesssim 245,000$ & \nodata & \nodata &$\gtrsim 240,000$\\
  \nv    &                        & $ 15.96\dagnote$ &  $\le 12.407 $       & & & \\
\hline	     	  	            	  	        	  		 	  
  \ovi   &  $2.440560\pm0.000014$ & $  9.76\pm3.10$ &  $13.370\pm    0.273$ & $\lesssim 91,000$ & \nodata & \nodata &$\gtrsim 230,000$\\
  \nv    &                        & $  9.76\dagnote$ &  $\le 12.450$      & & & \\
\hline	     	  	            	  	        	  		 	  
  \ovi   &  $2.440737\pm0.000147$ & $ 26.79\pm9.20$ &  $13.559\pm    0.262$ & $\le 690,000$ & \nodata & \nodata & $\sim 224,000$\\
  \nv    &                        & $ 26.79\dagnote$ &  $12.828\pm    0.081$ & & & \\
\hline	     	  	            	  	        	  		 	  
  \ovi   &  $2.437790\pm0.000062$ & $ 23.58\pm5.75$ &  $13.238\pm    0.126$ & $\le 535,000$ & \nodata & \nodata & $\sim 230,000$\\
  \nv    &                        & $ 23.58\dagnote$ &  $12.389\pm    0.068$ & & &  \\
\hline	     	  	            	  	        	  		 	  
  \ovi   &  $2.438248\pm0.000040$ & $ 20.14\pm5.46$ &  $13.250\pm    0.158$ & $\le 390,000$ & \nodata & \nodata & $\sim 240,000$\\
  \nv    &                        & $ 20.14\dagnote$ &  $12.185\pm    0.077$ & & & \\
\hline	     	  	            	  	        	  		 	  
  \ovi   &  $2.438963\pm0.000020$ & $ 36.57\pm3.59$ &  $13.827\pm    0.035$ & $<1,285,000$ & \nodata & \nodata & $\sim 270,000$\\
  \nv    &                        & $ 36.57\dagnote$ &  $12.338\pm    0.162$ & & & \\
\hline	     	  	            	  	        	  		 	  
  \civ   &  $2.439829\pm0.000041$ & $ 50.00\pm6.16$ &  $12.893\pm    0.042$ & & & \\
  \civ   &  $2.440775\pm0.000014$ & $ 12.50\pm1.55$ &  $12.782\pm    0.048$ & & & \\
  \siiii &  $2.437852\pm0.000024$ & $ 10.97\pm3.30$ &  $11.401\pm    0.117$ & & & \\
  \siiii &  $2.438954\pm0.000039$ & $ 17.29\pm2.60$ &  $11.934\pm    0.110$ & & &   \\
\hline
\hline
\multicolumn{8}{c}{Voigt Profile Components: $z=2.578$ absorption system}\\
\hline
\hline
  \civ   & $2.578167\pm0.000007$ &  $25.64\pm0.97$  & $13.340\pm0.014$ & $\le 350,000$ & $\ge 13.26$ & 13.332 & 27,600 \\ 
  \ciii  &                       &  $25.64\dagnote$ & $13.275\pm0.018$ & & & 13.270 \\ 
  \siiv  &                       &  $19.55\pm12.37$ & $11.671\pm0.213$ & & & 11.685 \\ 
  \hi    &                       &  $29.06\pm0.22$  & $\lesssim15.3$   & & & 14.276 \\
\hline
  \civ   & $2.578870\pm0.000004$ &  $22.30\pm0.67$  & $13.177\pm0.014$ & $\le 350,000$ & \nodata & 13.112 & 26,500 \\ 
  \ciii  &                       &  $22.30\dagnote$ & $13.458\pm0.014$ & & & 13.457 \\ 
  \siiv  &                       &  $14.58\dagnote$ & $11.883\pm0.091$ & & & 11.960 \\ 
  \hi    &                       &  $29.06\pm0.22$  & $15.642\pm0.014$ & & & 15.485 \\ 
\hline
  \ovi   & $2.578121\pm0.000054$ &  $22.86\pm2.47$  & $13.669\pm0.223$ & $\le 500,000$ & \nodata & \nodata & $\gtrsim 230,000$ \\ 
  \nv    &                       &  $24.43\dagnote$ & $12.749\pm0.186$ & & & \\ 
\hline
  \ovi   & $2.578495\pm0.000071$ &  $28.90\pm4.29$  & $13.930\pm0.124$ & $\le 800,000$ & \nodata & \nodata & $\gtrsim 250,000$ \\ 
  \nv    &                       &  $30.89\dagnote$ & $12.851\pm0.158$ & & & \\ 
\hline
  \ovi   & $2.579176\pm0.000022$ &  $9.51\pm3.05$  & $12.640\pm0.115$ & & & \\ 
  \hi    & $2.575657\pm0.000026$ &  $ 8.20\pm4.58$ & $11.718\pm0.254$ & & & \\ 
  \hi    & $2.575862\pm0.000034$ &  $34.67\pm3.04$ & $12.688\pm0.037$ & & & \\ 
  \hi    & $2.577126\pm0.000028$ &  $38.66\pm2.07$ & $13.560\pm0.033$ & & & \\ 
  \hi    & $2.577974\pm0.000005$ &  $26.29\pm0.33$ & $15.568\pm0.014$ & & & \\ 
  \hi    & $2.580563\pm0.000006$ &  $33.14\pm0.65$ & $13.198\pm0.007$ & & & \\ 

\enddata

\tablenotetext{\dagnote~}{\small Denotes $b$ parameters which have
been tied at the ratio appropriate for thermal broadening (relative to
other component ions).}

\tablenotetext{a}{\small The \hi $b$ parameter for this component
cannot be measured from the \hi profile itself.  The inferred value
from the relative \nv, \civ, and \siiv linewidths is highly uncertain,
but is probably at least $\gtrsim 25$ km/s, with an unknown (but
non-zero) contribution from turbulence or bulk gas motions.}

\tablenotetext{b}{\small \hi profile is saturated.  $b$ parameter and
column density for \hi are chosen for consistency with FOS spectrum in
\citep{HS1700_FOS} (See text).  }

\end{longtable}

\label{tbl:vpfit}
\clearpage
\begin{deluxetable}{c | l l l | l l l c }
\tablewidth{0pc}
\tablecaption{Best-Fit CLOUDY Model Parameters}
\tablehead{ & \multicolumn{3}{c}{{\em Fit Parameters}} & \multicolumn{3}{c}{{\em Derived Parameters}} & \\
{$z$} & {[X/H]} & {$\log(n_H)$} & {[Si/C]} & {$\Delta L$ (pc)\tablenotemark{1}} & {$\log(\nhi$)} & {$T$} (K) & {Line Constraints} }
\tablecolumns{7}
\startdata

1.844840 & $0.00^{+0.5}_{-0.5}$   & $-2.60^{+0.1}_{-0.1}$   & $-0.3^{+0.2}_{-0.2}$ & $64^{+113}_{-48}$ & $14.641^{+0.4}_{-0.6}$ & $9500^{+6500}_{-700}$   & {\scriptsize $\mhi\dagnote, \mcii, \mciv, \msiii\dagnote,\msiiii, \msiiv$} \\
1.845035 & $0.00^{+0.5}_{-0.5}$   & $-2.25^{+0.15}_{-0.10}$ & $-0.2^{+0.1}_{-0.1}$ & $104^{+63}_{-83}$ & $15.581^{+0.5}_{-0.5}$ & $8100^{+7000}_{-200}$   & {\scriptsize $\mhi\dagnote, \mcii, \msiii, \msiiii, \msiiv, \mmgii, \malii$ }\\
1.845044 & $0.00^{+0.5}_{-1.0}$   & $-3.20^{+0.2}_{-0.1}$   & \nodata              & $1208^{+12000}_{-850}$  & $14.619^{+1.1}_{-0.5}$ & $13,800^{+13,000}_{-1000}$ & {\scriptsize $\mhi\dagnote,\mnv, \mciv, \msiiv$} \\
1.845410 & $0.00\tablenotemark{a}$ & $-2.70^{+0.2}_{-0.2}$  & $-0.4^{+0.5}_{-0.5}$ & $96^{+100}_{-60}$  &   $14.607^{+0.6}_{-0.6}$ & $10,100^{+1000}_{-500}$ & {\scriptsize $\mciv, \msiiii, \msiiv$} \\
1.845511 & $0.00\tablenotemark{a}$ & $-2.30^{+0.3}_{-0.3}$  & $-0.4^{+0.4}_{-0.5}$ & $24^{+100}_{-10}$ & $14.844^{+0.7}_{-0.4}$  & $8400^{+1100}_{-800}$ & {\scriptsize $\mciv, \msiiii, \msiiv$} \\

\hline
2.167935 & $-2.00^{+1.5}_{-0.5}$   & $-2.95^{+0.35}_{-0.0}$  & $0.6^{+0.1}_{-0.4}$ & $39109^{+17052}_{-38582}$  & $16.368^{+0.37}_{-1.2}$  & $29,900^{+0}_{-10,000}$   & {\scriptsize $\mhi\dagnote, \mcii, \mciv, \msiiii, \msiiv, \mmgii, \malii$} \\
2.168064 & $-2.25^{+0.75}_{-0.25}$  & $-2.00^{+0.1}_{-0.35}$ & $0.6^{+0.1}_{-0.3}$ & $898^{+598}_{-520}$  & $16.246^{+0.3}_{-0.6}$  & $22,000^{+3500}_{-2000}$   & {\scriptsize $\mhi\dagnote, \mcii, \mciv, \msiiii, \msiiv, \mmgii, \malii$} \\

\hline
2.314470 & $0.00^{+0.5}_{-1.2}$ &$-3.00^{+0.35}_{-0}$& \nodata             & $32^{+475}_{-24}$      & $13.127^{+1.60}_{-0.45}$ & $11,000^{+12,700}_{-1000}$ & {\scriptsize $\mhi\dagnote, \mciv, \msiiv$} \\
2.314895 & $0.00^{+0.25}_{-0.25}$  & $-1.50^{+0.30}_{-0.40}$ & $0.0^{+0.1}_{-0.1}$ & $2.0^{+1.0}_{-1.8}$    & $15.084^{+0.25}_{-0.01}$ & $5400^{+6500}_{-400}$ & {\scriptsize $\mnii ,\malii ,\mmgii ,\mcii ,\msiii $ } \\
2.314996 & $-0.25^{+0.25}_{-0.75}$ & $-1.00^{+0.30}_{-0.25}$ & $0.3^{+0.2}_{-0.3}$ & $0.7^{+1.9}_{-0.1}$   & $15.576^{+0.70}_{-0.25}$ & $\lesssim5400$ & {\scriptsize $\mhi\dagnote,\mnii ,\malii ,\mmgii ,\mcii ,\msiii ,\mfeii $ } \\
2.315504 & $0.00^{+0.5}_{-0.5}$    & $-1.00^{+0.15}_{-0.0}$  & $0.0^{+0.1}_{-0.1}$ & $0.6^{+0.68}_{-0.40}$ & $15.569^{+0.544}_{-0.459}$ & $\lesssim5400$ & {\scriptsize $\mhi\dagnote,\mnii ,\malii ,\mmgii ,\mcii ,\msiii ,\mfeii $ } \\
2.316350 & $0.25^{+0.25}_{-0.5}$   & $-2.90^{+0.00}_{-0.2}$  & \nodata             & $292^{+773}_{-99} $ & $14.128^{+0.328}_{-0.428}$ & $18,000^{+1000}_{-8000}$ & {\scriptsize $\mhi\dagnote, \mciv, \msiiv,\mnv$} \\
2.314992 &        \nodata          & \nodata              & \nodata & \nodata        &       \nodata     & $\gtrsim 230,000$     &  {\scriptsize $\movi, \mnv\dagnote$}\\ 
2.315182 &        \nodata          & \nodata              & \nodata & \nodata        &       \nodata     & $\gtrsim 250,000$     &  {\scriptsize $\movi, \mnv\dagnote$}\\ 
2.315790 &        \nodata          & \nodata              & \nodata & \nodata        &       \nodata     & $\gtrsim 250,000$     &  {\scriptsize $\movi, \mnv\dagnote$}\\ 
2.316345 &        \nodata          & \nodata              & \nodata & \nodata        &       \nodata     & $\gtrsim 230,000$     &  {\scriptsize $\movi, \mnv\dagnote$}\\ 
2.316693 &        \nodata          & \nodata              & \nodata & \nodata        &       \nodata     & $\gtrsim 260,000$     &  {\scriptsize $\movi, \mnv\dagnote$}\\ 
2.314618 &        \nodata          & \nodata              & \nodata & \nodata        &       \nodata     & $\gtrsim 260,000$     &  {\scriptsize $\movi, \mnv\dagnote$}\\ 
2.314900 & $0.00^{+0.5}_{-0.25}$   & $-2.60^{+0.25}_{-0.0}$  & \nodata & $405^{+230}_{-297}$ & $15.094^{+0.321}_{-0.360}$ & $8300^{+7200}_{-1300}$ & {\scriptsize $\mciv, \mnv, \msiiv, \mhi\dagnote$}\\ 
2.315117 & $0.00^{+0.5}_{-2}$ & $-3.10^{+0.15}_{-0.0}$& \nodata & $784^{+113,000}_{-545}$& $14.286^{+2}_{-0.567}$& $12,000^{+22,000}_{-1000}$ & {\scriptsize $\mciv, \mnv, \msiiv\dagnote, \mhi\dagnote$}\\ 
2.316071 & $0.00^{+0.5}_{-1.0}$    & $-3.20^{+0.2}_{-0.2}$   & \nodata & $130^{+1235}_{-89}$ & $13.290^{+1.213}_{-0.473}$ & $12,500^{+15,000}_{-1500}$ & {\scriptsize $\mciv, \mnv, \msiiv, \mhi\dagnote$}\\ 
2.315868 & $-0.25^{+0.75}_{-0.75}$ & $-3.15^{+0.15}_{-0.05}$ & \nodata & $574^{+2842}_{-470}$& $13.972^{+0.929}_{-0.748}$ & $15,500^{+10,000}_{-5500}$ & {\scriptsize $\mciv, \mnv, \msiiv, \mhi\dagnote$}\\ 
2.315556 & $-0.25^{+0.75}_{-1.0}$  & $-3.05^{+0.1}_{-0.05}$ & \nodata & $1280^{+17,600}_{-1048}$& $14.536^{+1.164}_{-0.756}$ & $14,000^{+10,000}_{-4000}$ & {\scriptsize $\mciv, \mnv, \msiiv\dagnote, \mhi\dagnote$}\\ 
2.313843 & $0.0^{+0.1}_{-0.1}$     & $-2.15^{+0.1}_{-0.1}$ & \nodata & $6^{+3}_{-3}$         & $14.214^{+0.100}_{-0.100}$ & $6600^{+6500}_{-300}$ & {\scriptsize $\mciv, \mhi$}\\ 

\hline
2.379919 & $-2.0^{+0.2}_{-0.2}$    & $-3.05^{+0.2}_{-0.3}$& $0.3^{+0.3}_{-0.3}$ & $5448^{+2000}_{-2000}$  & $15.294^{+0.2}_{-0.1}$ & $29,000^{+2000}_{-2000}$ & {\scriptsize $\mhi, \mciii, \mciv, \msiiii, \msiiv, \msiii\dagnote$}\\
2.379988 & \nodata                 & \nodata              & & \nodata                 & \nodata                & $\gtrsim 280,000\tablenotemark{*}$ & {\scriptsize $\movi, \mnv\tablenotemark{$\dagger$}$} \\

\hline
2.431814 & $-2.75^{+0.25}_{-0.25}$ & $-2.85^{+0.2}_{-0.3} $  & \nodata                & $2659^{+10,350}_{-1850}$& $14.861^{+0.144}_{-0.070}$ & $35,000^{+3000}_{-3000}$ & {\scriptsize $\mhi, \mciii, \mciv$} \\
2.432091 & $-2.5^{+0.25}_{-0.25}$  & $-2.75^{+0.2}_{-0.25}$  & \nodata                & $5105^{+10,000}_{-3000}$& $15.370^{+0.150}_{-0.163}$ & $31,800^{+3000}_{-2100}$  & {\scriptsize $\mhi, \mciii, \mciv$} \\
2.432933 & $-1.75^{+0.75}_{-0.5}$ & $-2.25^{+0.25}_{-0.45}$ & \nodata                & $450^{+2203}_{-354}$    & $15.423^{+0.319}_{-0.437}$&  $24,000^{+7800}_{-7000}$ & {\scriptsize $\mhi, \mcii, \mciii, \mciv$}\\
2.433052 & $-1.5^{+0.5}_{-0.25}$   & $-1.90^{+0.20}_{-0.35}$ & $0.37^{+0.15}_{-0.15}$ & $232^{+433}_{-110}$   & $15.888^{+0.38}_{-0.38}$&  $19,700^{+3300}_{-4700}$ & {\scriptsize $\mhi, \mcii, \mciii, \mciv, \msiii, \msiiii, \msiiv, \malii$}\\
2.433426 & $-1.5^{+0.5}_{-0.25}$   & $-1.90^{+0.30}_{-0.35}$ & $0.25^{+0.30}_{-0.25}$ & $218^{+460}_{-87}$    & $15.859^{+0.15}_{-0.42}$ & $19,700^{+3300}_{-4700}$  & {\scriptsize $\mhi, \mcii, \mciv, \msiii, \msiiii, \msiiv$ }\\
2.433650 & $-1.5^{+0.3}_{-0.5}$    & $-1.90^{+0.1}_{-0.45}$  & $0.30^{+0.2}_{-0.3}$   & $112^{+780}_{-20}$    & $15.570^{+0.32}_{-0.24}$ & $19,700^{+4800}_{-2700}$  & {\scriptsize $\mhi\tablenotemark{$\dagger$}, \mcii, \mciv, \msiiii, \msiiv, \malii$ }\\
2.432600 & \nodata                 & \nodata                 & \nodata                & \nodata               & \nodata                  & $\gtrsim 224,000$ & {\scriptsize $\movi, \mnv\tablenotemark{$\dagger$}$} \\
2.433233 & \nodata                 & \nodata                 & \nodata                & \nodata               & \nodata                  & $\gtrsim 280,000$ & {\scriptsize $\movi, \mnv\tablenotemark{$\dagger$}$} \\
2.433749 & \nodata                 & \nodata                 & \nodata                & \nodata               & \nodata                  & $\gtrsim 280,000$ & {\scriptsize $\movi, \mnv\tablenotemark{$\dagger$}$} \\
2.437774 & $-1.25^{+0.25}_{-0.25}$& $-2.75^{+0.1}_{-0.2}$ & $0.4^{+0.5}_{-0.4}$  & $168^{+630}_{-14}$ & $13.961^{+0.121}_{-0.061}$ &  $27,000^{+3000}_{-6000}$  & {\scriptsize $\mhi, \mciii, \mciv, \msiiii, \msiiv$} \\
2.438286 & $-2.25^{+0.25}_{-0.25}$& $-2.65^{+0.10}_{-0.25}$& $0.20^{+0.1}_{-0.3}$& $1596^{+1655}_{-50}$ & $15.086^{+0.363}_{-0.125}$  & $31,800^{+2300}_{-7800}$ & {\scriptsize $\mhi, \mcii, \mciii, \mciv, \msiii, \msiiii, \msiiv $}\\
2.438637 & $-2.0^{+0.25}_{-0.25}$& $-2.90^{+0.5}_{-0.05}$& $0.60^{+0.2}_{-0.25}$ & $4413^{+327}_{-3758}$ & $15.510^{+0.162}_{-0.098}$  & $28,500^{+8300}_{-6000}$ & {\scriptsize $\mhi, \mcii, \mciii, \mciv, \msiii, \msiiii, \msiiv $}\\
2.438857 & $-1.5^{+0.25}_{-0.5}$ & $-2.20^{+0.15}_{-0.5}$& $0.20^{+0.3}_{-0.2}$  & $225^{+1850}_{-75}$ & $15.240^{+0.26}_{-0.24}$  &  $22,000^{+8000}_{-3500}$  & {\scriptsize $\mhi, \mcii, \mciii, \mciv, \msiii, \msiiii, \msiiv $}\\
2.440517 & $-1.5^{+0.25}_{-0.5}$ & $-2.60^{+0.2}_{-0.7}$ & $0.0^{+0.8}_{-0.0}$   & $323^{+7500}_{-198}$ & $14.536^{+0.100}_{-0.100}$ &  $27,000^{+12,000}_{-4800}$ & {\scriptsize $\mhi, \mciv, \msiiv\tablenotemark{$\dagger$}$} \\
2.439748 & \nodata                 & \nodata                 & \nodata                & \nodata               & \nodata                  & $\gtrsim 234,000$ & {\scriptsize $\movi, \mnv\tablenotemark{$\dagger$}$} \\
2.440253 & \nodata                 & \nodata                 & \nodata                & \nodata               & \nodata                  & $\gtrsim 240,000$ & {\scriptsize $\movi, \mnv\tablenotemark{$\dagger$}$} \\
2.440560 & \nodata                 & \nodata                 & \nodata                & \nodata               & \nodata                  & $\gtrsim 230,000$ & {\scriptsize $\movi, \mnv\tablenotemark{$\dagger$}$} \\
2.440737 & \nodata                 & \nodata                 & \nodata                & \nodata               & \nodata                  & $\gtrsim 224,000$ & {\scriptsize $\movi, \mnv\tablenotemark{$\dagger$}$} \\
2.437790 & \nodata                 & \nodata                 & \nodata                & \nodata               & \nodata                  & $\gtrsim 230,000$ & {\scriptsize $\movi, \mnv\tablenotemark{$\dagger$}$} \\
2.438248 & \nodata                 & \nodata                 & \nodata                & \nodata               & \nodata                  & $\gtrsim 240,000$ & {\scriptsize $\movi, \mnv\tablenotemark{$\dagger$}$} \\
2.438963 & \nodata                 & \nodata                 & \nodata                & \nodata               & \nodata                  & $\gtrsim 270,000$ & {\scriptsize $\movi, \mnv\tablenotemark{$\dagger$}$} \\

\hline
2.578121 & \nodata                 & \nodata               & & \nodata                 & \nodata                 & $\gtrsim 230,000\tablenotemark{*}$ & {\scriptsize $\movi, \mnv$} \\
2.578167 & $-1.0^{+0.75}_{-1.0}$   & $-3.35^{+0.15}_{-0.1}$& $0.4^{+0.4}_{-0.4}$ & $1878^{+35,000}_{-1700}$  & $14.276^{+1.0}_{-0.6}$& $27,600^{+12,000}_{-10,000}$ & {\scriptsize $\mhi\dagnote, \mciv, \mciii, \msiiv$} \\
2.578495 & \nodata                 & \nodata               & & \nodata                 & \nodata                 & $\gtrsim 250,000\tablenotemark{*}$ & {\scriptsize $\movi, \mnv$} \\
2.578870 & $-2.0^{+0.1}_{-0.1}$    & $-2.95^{+0.1}_{-0.1}$ & $0.3^{+0.2}_{-0.3}$ & $5122^{+1000}_{-1000}$  & $15.485^{+0.15}_{-0.1}$  & $26,500^{+2000}_{-2000}$ & {\scriptsize $\mhi, \mciv, \mciii, \msiiv$} \\

\enddata

\tablenotetext{1}{\small Line of sight distance through absorbing shell, averaged for all heavy element line measurements.  Uses best-fit [X/H], $\log n_H$, and the CLOUDY ionization fractions that correspond to these parameters.}
\tablenotetext{$\dagger$}{\small Upper limit}
\tablenotetext{*}{\small Alternate solution is available with $L\lesssim 100$ kpc, $\log n_H \sim -4$, $[X/H]\gtrsim -1.5$, photoionized.}
\tablenotetext{a}{\small Denotes systems where an accurate metallicity could not be measured because of confusion and/or blending in the \hi profile.  
For these cases, we assume the metallicity matches other, measured components in the same complex with similar heavy element ion ratios.}

\end{deluxetable}

\label{tbl:models}

\end{document}